\documentclass[manuscript]{aastex}          
\usepackage{spr-astr-addons}    
\usepackage{natbib}
\usepackage{epstopdf}
\usepackage{graphicx}
\usepackage{bm}
\usepackage[breaklinks=true,colorlinks=true,linkcolor=blue,urlcolor=blue,citecolor=blue]{hyperref}

\newcommand{\Msun}{$M_\odot$}
\newcommand{\Ls}{L_{\rm s}^\infty}
\newcommand{\Lh}{L_{\rm h}^\infty}

\newcommand{\Tg}{\widetilde{T}}
\newcommand{\Ts}{T_{\rm s}^\infty}

\newcommand{\gcc}{\mathrm{g~cm}^{-3}}
\newcommand{\rhoacc}{\rho_{\rm acc}}

\begin{document}

\title{Afterburst thermal relaxation in neutron star crusts}

\shorttitle{Afterburst relaxation in neutron star crusts}

\shortauthors{Chaikin et al.}

\author{E. A. Chaikin\altaffilmark{1}}
\and
\author{A. D. Kaminker\altaffilmark{2}}
\and
\author{D. G. Yakovlev\\
\altaffilmark{2,*}}
\altaffiltext{*}{yak.astro@mail.ioffe.ru}
\email{yak.astro@mail.ioffe.ru}

\altaffiltext{1}{Peter the Great St~Petersburg Polytechnic University, Politekhnicheskaya 29, 194021, St~Petersburg, Russia. Current address: Argelander-Institut f\"ur Astronomie, Auf dem H\"ugel 71, D-53121, Bonn, Germany}

\altaffiltext{2}{Ioffe Institute, Politekhnicheskaya
	26, 194021, St Petersburg, Russia}

\newcommand{\vdag}{(v)^\dagger}
\newcommand{\myemail}{yak.astro@mail.ioffe.ru}

\begin{abstract}
We study thermal relaxation in a neutron star after 
internal heating events (outbursts) in the crust. 
We consider thin and thick spherically symmetric heaters,
superfluid and non-superfluid crusts, stars with open and forbidden direct
Urca processes in their cores. In
particular, we analyze long-term thermal relaxation after
deep crustal heating produced by nuclear transformations in
fully or partly accreted crusts of transiently accreting neutron stars. This  
long-term relaxation has a typical relaxation time and an overall finite
duration time for the crust to thermally equilibrate with the core. Neutron superfluidity
in the inner crust greatly affects the relaxation if the heater is located in the inner
crust. It shortens and unifies the time of emergence of thermal wave from the
heater to the surface. This is important for
the interpretation of observed outbursts of magnetars and
transiently accreting neutron stars in quasi-persistent low-mass
X-ray binaries.
\end{abstract}






\section{Introduction}
\label{sec:level1}

Neutron stars are born hot in supernova explosions but then
gradually cool down loosing their internal thermal energy via
neutrino emission from their interiors and via the thermal surface
radiation, as reviewed, e.g., by \citet{YakPeth2004,2006PGW,
PPP2015}. A study of cooling neutron stars allows one to explore
still poorly known properties of dense matter in neutron star
interiors. For instance, investigation of middle-aged isolated
ordinary neutron stars may give valuable information on neutrino
emission mechanisms and superfluid properties of neutron star cores
-- bulky internal regions containing about 99 per cent of the
neutron star mass (e.g., \citealt{HPY2007}). Such stars are
isothermal inside because of high internal thermal conduction, with
substantial temperature gradients persisting only in the so called
heat blanketing envelopes near their surface, not wider than a few
hundred meters; e.g., \citet{PPP2015}. Owing to a high thermal
conductivity, the stellar core is thermally coupled to the crust, so
that the thermal luminosity reflects physical conditions in the
core.

The situation is more complicated if neutron stars are
reheated from inside. The reheating mechanisms have been reviewed, e.g., by 
\citealt{YakPeth2004,2006PGW}, and in references
therein. These mechanisms (for instance, reheating by
differential rotation or ohmic decay of magnetic fields) can operate
in different places of the star, in its core and/or crust. The
deeper the heating, the larger fraction of the heat is emitted by
neutrinos, and the weaker the thermal surface luminosity
(e.g., \citealt{2006Kam,2014HEAT} and references therein).

In this paper, we continue to study the effects of the 
heaters, which produce outbursts in the crust, 
and subsequent relaxation of the surface thermal emission. The aim is to elaborate
the methods of inferring the properties of the heaters from
observations of the surface emission. In particular, we expect that
our results will be useful for the interpretation of observations of
old transiently accreting neutron stars in low-mass X-ray binaries
(LMXBs) and warm active magnetars. As suggested by \citet{1990HZ,1998BBR}, 
neutron stars in LMXBs can be reheated by nuclear transformations (deep crustal heating processes) 
in the accreted crust. This mechanism has been
elaborated later (e.g.,
\citealt{2007Shternin,2009BC,2012PR,2013PR,2015TAP,
2014MC,2014HOMAN,2016Merritt,2018Aql}). Magnetars, which are mainly hot,
middle-aged and active neutron stars (reviewed, e.g. by
\citealt{MAGNETARS,2017KB}) are most probably reheated by the
processes related to strong magnetic fields (e.g.
\citealt{2009Pons,2011PonsPerna,2012PonsRea,2013Vigano,2016BL,2016LLB,2018MagnetarOutburs} and references therein).

We have selected the crustal heaters because they are closer to the
stellar surface; it is easier for them to affect the thermal surface
emission. Such heaters can break isothermality of
neutron star interiors and violate then the thermal
coupling between the core and the surface but establish the thermal
coupling between the heater and the surface. In the latter case, the
star becomes a natural laboratory of its heater. In the first part
of this paper
we do not specify the nature of the heater
but treat it in a phenomenological way. For
quasi-stationary heaters, we have done it earlier \citep{2006Kam,2007Kam,2009HEAT,2014HEAT}. Recently, we
have considered variable heaters in a non-superfluid crust
\citep{2017Ch}. Now we study stars with neutron
superfluidity in the inner crust which is known to be
important (e.g. \citealt{2009BC}). Preliminary results
of these studies have been reported by \citet{JPCS}. 

In Sections \ref{sec:level2} and \ref{sec:level3} we outline
microphysics input and our cooling code. In Section \ref{sec:level4}
we model the heaters operating in thin spherical
layers. In Sections \ref{sec:level6} and \ref{sec:relax} we apply
the theory for the interpretation of the long-term relaxation in neutron
stars entering quasi-persistent LMXBs. Conclusions are formulated in Section
\ref{sec:level5}.

\section{Numerical simulations}
\label{sec:level2}

To simulate thermal evolution of a neutron star we use our
one-dimensional code written in Python programming language. As in
\citet{2017Ch,JPCS}, we consider a spherically symmetric neutron star
including the effects of General Relativity. We use a mesh of 350
radial spherical cells from the stellar center to the density $\rho
= \rho_{\rm b} \sim 10^9-10^{10}$ g cm$^{-3}$ at the bottom of the
heat blanketing envelope (e.g., \citealt{1983GPE,1997POT,2016BPY});
$\rho_{\rm b}$ can be varied depending on the problem (e.g.,
\citealt{2016BPY}). For the densities $\rho<\rho_{\rm b}$, we  use
separately calculated $T_{\rm b}-T_{\rm s}$ relations, where $T_{\rm
b}$ is the temperature at $\rho=\rho_{\rm b}$ and $T_{\rm s}$ is the
local effective temperature of the stellar surface.
Our code is based on the implicit Euler backward method, 
which easily handles high temperature gradients and stabilizes simulations
under abrupt releases of heater's energy in the star.

We consider two neutron star models with the BSk21 equation of
state of nucleon matter in the core (e.g.
\citealt{BSk2013}). The gravitational masses of these stars are $M = 1.4$ 
and $1.85\,{M}_\odot$, their circumferential radii are $R=12.6$ and
$12.5$ km, and their surface gravities are
$g_{\rm s}=1.43 \times 10^{14}$ and $2.11 \times 10^{14}$ cm~s$^{-2}$, respectively. 

For simplicity, we neglect the effects of nucleon
superfluidity in the star's core but include them in the
crust where the heaters are located. The $1.4\,
{ M}_\odot$ star is an example of the standard neutrino
candle whose neutrino cooling is mainly provided by the modified
Urca process of neutrino emission from the star's core (e.g.,
\citealt{Yak2001} and references therein). The $1.85\,
{M}_\odot$ star has a central kernel where the powerful
direct Urca neutrino process is allowed \citep{1991DURCA}; it
undergoes fast neutrino cooling. While
studying transient thermal processes in a crust one
often considers the effect of ion impurities on the
electron thermal conductivity (e.g., 
\citealt{2012PR}). We neglect the impurities here to
avoid introducing largely unknown impurity distribution.

We treat the heater as a spherical layer where a given heat power
$Q(\rho,t)$ [erg s$^{-1}$ cm$^{-3}$] is released. The total redshifted heat power
$\Lh(t)$ [erg s$^{-1}$] is
\begin{equation}
\Lh(t) = \int Q(\rho, t)\, \exp(2\Phi) \, {\rm d}V, \label{eq:LLL}
\end{equation}
where d$V$ is a proper volume element and $\Phi$ is the metric
function that determines gravitational redshift (e.g.,
\citealt{HPY2007}). The cooling code calculates the redshifted
effective surface temperature $\Ts(t)$ and
redshifted surface thermal luminosity $\Ls(t)$. 
The problem is whether the $\Ls(t)$ variations  are observable. 

\section{Superfluidity}
\label{sec:level3}

Earlier we \citep{2017Ch} have studied this problem neglecting
superfluidity. However, neutron stars are thought to be
superfluid inside. A comprehensive review of superfluidity is
given by \citet{2014SF}. In our case, free neutrons can be
superfluid due to the singlet-state ($^1S_0$) Cooper pairing in the
inner crust (at densities $\rho$ higher than the
neutron drip density $\rho_{\rm drip} \sim (4 - 6) \times 10^{11}$ g~cm$^{-3}$). 
Also, neutrons can be superfluid owing to the triplet-state
pairing, and protons owing to the singlet-state pairing in the
star's core. Each superfluidity is characterized by
appropriate critical temperature $T_{\rm c}(\rho)$ which depends on
pairing particles and pairing type as well as on $\rho$.
A study of $T_{\rm c}(\rho)$ is a fundamental problem of
neutron star physics. Calculations of $T_{\rm c}(\rho)$ are complicated
because of many-body (polarization) effects. Various versions of
many-body theories lead to vastly different $T_{\rm c}(\rho)$. In
this situation, it is instructive to try different theoretical
versions. As mentioned above, we
assume a non-superfluid core and study the difference
in the $\Ls(t)$ behavior between superfluid and non-superfluid
neutrons in the inner crust.

\begin{figure}
\centering
\includegraphics[width=0.98\columnwidth]{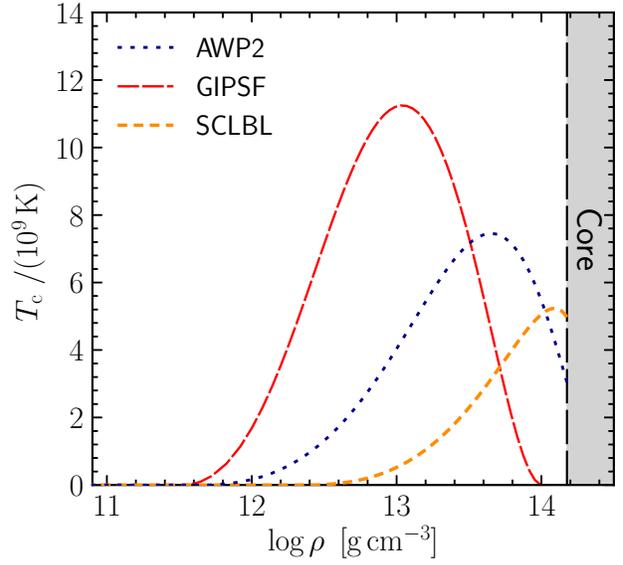}  
\caption{\label{fig:Tcrit}Critical temperature $T_{\rm c}$ for
singlet-state pairing of free neutrons in a neutron star crust as a
function of $\rho$ for three models of neutron superfluidity (AWP2,
GIPSF and SCLBL). The long-dashed vertical line
shows the position of the crust-core interface ($\rho = \rho_{\rm
cc} \sim 1.5 \times 10^{14}\ \gcc$). Calculations include the
effects of superfluidity only in the inner crust.}
\end{figure}

We consider three models of $T_{\rm c}(\rho)$ in the crust (Fig.\
\ref{fig:Tcrit}). The GIPSF
(Gandolfi--Illarionov--Pederiva--Schmidt--Fantoni; 
\citealt{2009GIPSF}) model  is close to the pure BCS
(Bardeen--Cooper--Schrieffer) model and gives  high and broad
$T_{\rm c}(\rho)$ peak. It can strongly affect $\Ls(t)$. The
AWP2 (Ainsworth--Wambach--Pines; \citealt{1989AWP}) model is
stronger affected by the in-medium polarization effects; it
weakens the influence of superfluidity on $\Ls(t)$. The SCLBL
(Schulze--Cugnon--Lejeune--Baldo--Lombardo; \citealt{1996SCLBL})
model is significant only if the heater is placed deeper in
the crust.

\begin{figure}
\centering
\includegraphics[width=0.98\columnwidth]{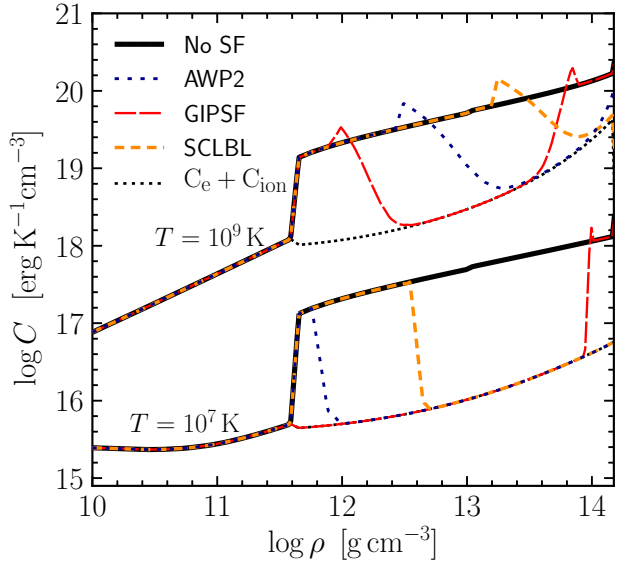} 
\caption{\label{fig:heatcap} Heat capacity $C$ per unit
volume vs. $\rho$
in a star's crust at two temperatures,
$T=10^7$ and $10^9$ K, for the three models of $^{1}S_{0}$
neutron superfluidity (AWP2, GIPSF and SCLBL). The thick solid lines
show $C$ in the absence of superfluidity; the lower dotted lines show
the heat capacity of electrons + ions.}
\end{figure}

The strongest effect of superfluidity in the crust is the effect on
the volumetric heat capacity of the matter, $C(\rho,T)$  (e.g.,
\citealt{1999REV,PPP2015}). We do not distinguish the heat
capacities at constant volume and pressure, because the difference 
is very small (e.g.,
\citealt{HPY2007}). In the absence of superfluidity, at 
$T > T_{\rm c}(\rho)$, the main contribution to the heat capacity
of the inner crust comes from free neutrons, but at $T\ll T_{\rm c}$
this contribution is
greatly suppressed by a gap in the
energy spectra of superfluid neutrons. Then the heat capacity is mostly
determined by much smaller contributions of ions (atomic nuclei) and
electrons (Fig.\ \ref{fig:heatcap}). The main processes of neutrino
emission in the crust are not affected by superfluidity of neutrons
\citep{Yak2001,PPP2015}. The thermal conductivity in the crust,
that is mainly provided by degenerate electrons, is
unaffected by neutron superfluidity as well (e.g.,
\citealt{PPP2015}). 

\section{Thin heating layer}
\label{sec:level4}

\begin{figure*}
\includegraphics[width=0.98\columnwidth]{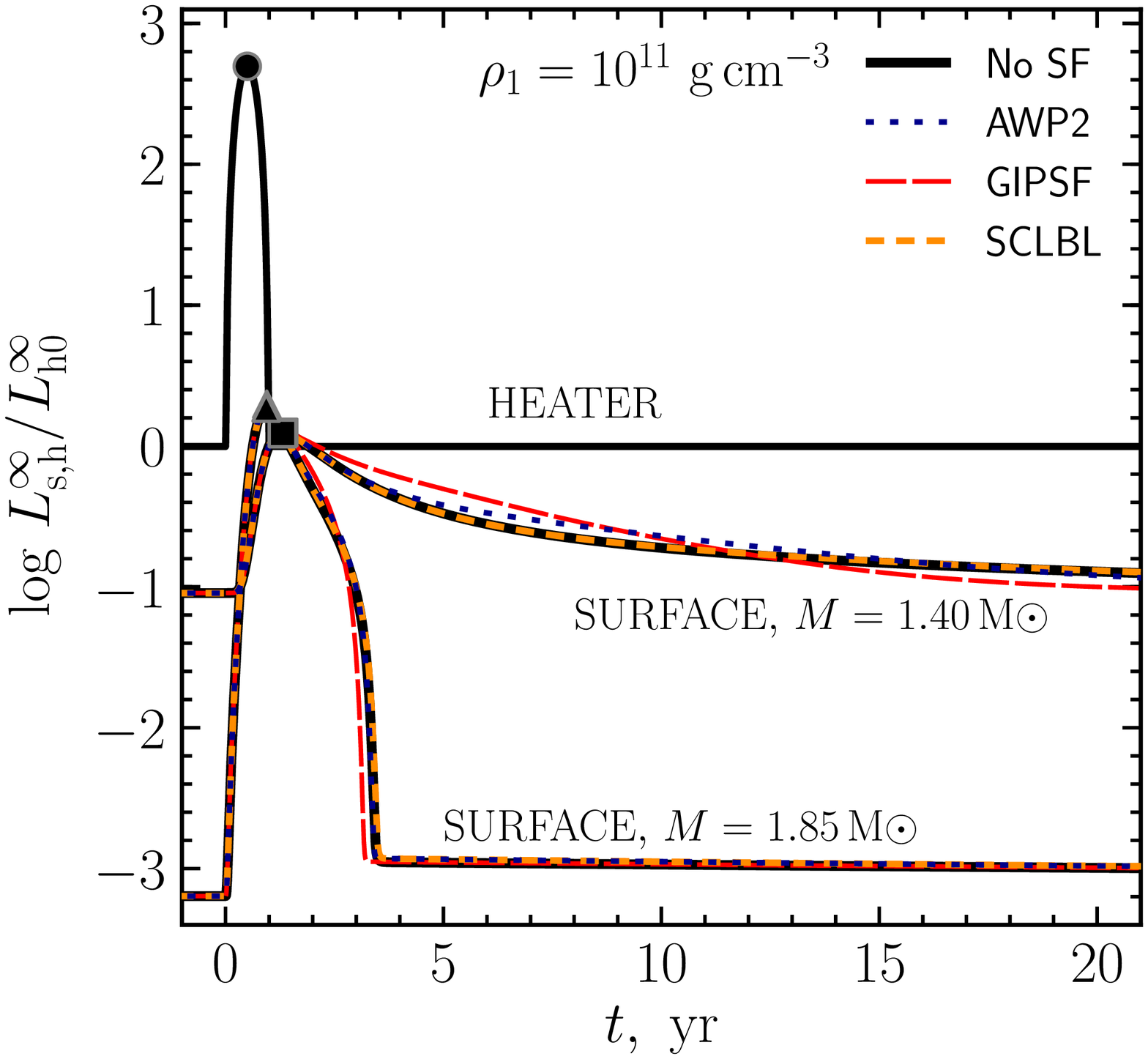}%
\includegraphics[width=0.99\columnwidth]{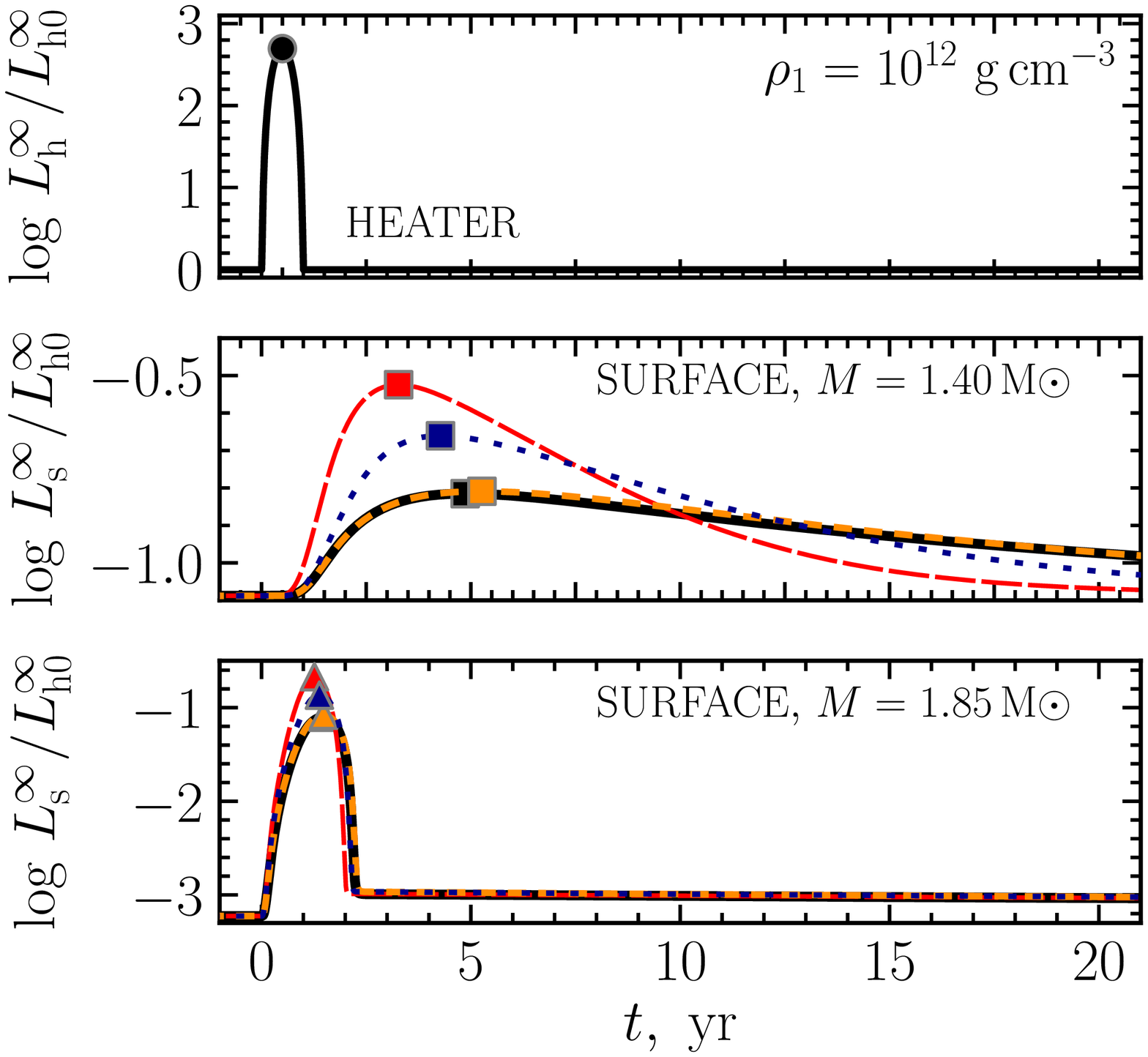}
\caption{\label{fig:main_cool} Time variations of the heater's power
$\Lh(t)$ (labeled as `HEATER', one and the same for a fixed $M$;
$\Delta t=1$ yr, $\alpha_{\rm h} = 500$) and
the surface luminosity $\Ls(t)$ (labeled as `SURFACE') in units of
$L_{\rm h0}^\infty$ for the 1.40 and 1.85\,\Msun\ 
stars with the three models of crustal superfluidity (AWP2, GIPSF
and SCLBL). The left panel corresponds to the heater position
$\rho_1 = 10^{11}$ $\gcc$. The
right panel is for $\rho_1 = 10^{12}$ $\gcc$. 
The solid curves $\Ls(t)$ display the
case of non-superfluid neutrons (`No SF'). The circles show maxima
of $\Lh(t)$; the squares and triangles position maxima of $\Ls(t)$
for the 1.40 and 1.85~\Msun\ stars, respectively.}
\end{figure*}

\begin{figure*}
\includegraphics[width=0.98\columnwidth]{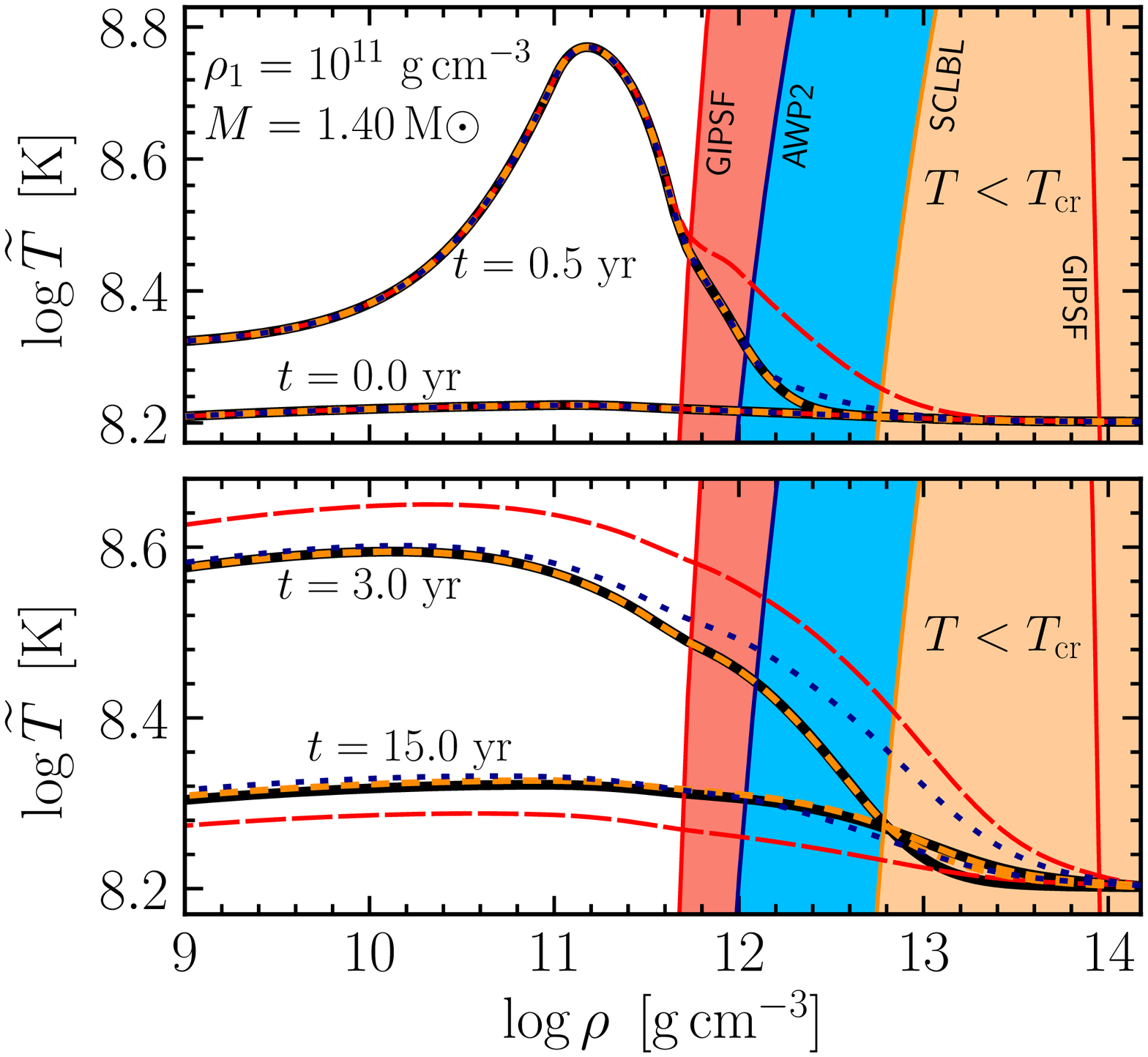}%
\includegraphics[width=0.98\columnwidth]{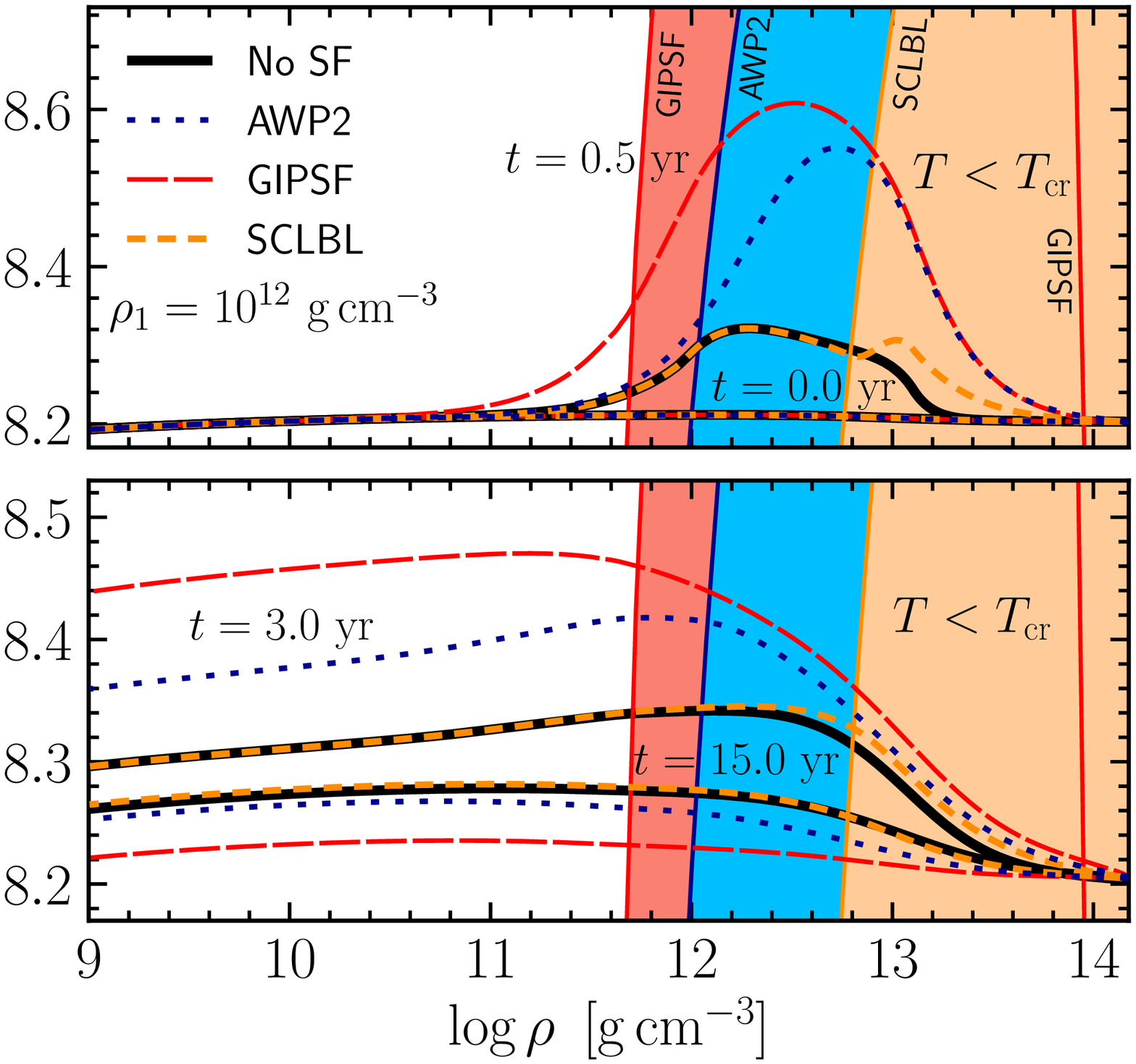}
\caption{\label{fig:main_profile} Temperature profiles $\Tg(\rho)$
within the crust at four moments of time $t$=0, 0.5, 3 and 15 yr from the beginning of a
heater's outburst for the 1.40\,\Msun\ star with the three models of
crustal superfluidity (Fig.\ \ref{fig:Tcrit}). The left and
right panels correspond to the same heater's parameters as in
Fig.\ \ref{fig:main_cool}. The solid curves refer to the case of
non-superfluid neutrons. The shaded regions show the areas of
superfluid neutrons, $T < T_{\rm c}(\rho)$. For GIPSF superfluidity,
the area is restricted by the two nearly vertical lines marked as
GIPSF. For AWP2 and SCLBL superfluidities, the areas are restricted
by the corresponding nearly vertical lines and the right vertical
axis (at $\rho=\rho_{\rm cc}$).}
\end{figure*}

Here we follow \cite{2017Ch} and treat the heater as a
thin spherical shell located at $\rho_1 < \rho < \rho_2$,
where $\rho_1$ 
and $\rho_2$ are some boundary values. Simulations are
performed in a few steps. First, an initially hot star cools freely
until thermal equilibrium is reached within the star. Then we turn on a stationary heater with a
time-independent heat power $Q(\rho, t) \equiv Q_{0}(\rho)$, $\Lh
\equiv L_{\rm h0}^\infty$, and wait until the
history of the initial thermal evolution is forgotten and a
stationary state within the star is established. After that we
generate the heater's outburst, $Q(\rho, t)$, and calculate $\Lh(t)$ as well as
the surface luminosity $\Ls(t)$.

The duration of surface 
outbursts is mainly determined by a
heat diffusion time $\tau_{\rm diff}$. According to
\cite{1969Henyey}, the diffusion time along the radial 
coordinate $r$ in an interval $r_1 \leq r \leq
r_2$ can be estimated as
\begin{equation}
   \tau_{\rm diff} \sim \frac{1}{4}
     \,\left[ \int_{r_1}^{r_2} {\rm d}r\,
     \left( \frac{C}{\kappa} \right)^{1/2} \right]^2,
    \label{eq:taudiff}
\end{equation}
where $\kappa$ and $C$ are, respectively, the thermal conductivity
and heat capacity per unit volume. Hence, $\tau_{\rm diff}$ 
is regulated by the $C/\kappa$ ratio.

In a star with an internal temperature $\sim 10^8$~K,
$C/\kappa$ mainly increases with growing $T$ and with the growth of
$\rho$ at $\rho \lesssim \rho_{\rm drip}$. However it may decrease with the
growth of $\rho$ at $\rho \gtrsim \rho_{\rm drip}$ under the
effect of crustal superfluidity; 
e.g., \cite{2009BC,2013PR,PPP2015} and references therein.

The relaxation times $\tau_{\rm diff}^{\rm
out}$ and $\tau_{\rm diff}^{\rm in}$ for the heat flows directed
 outside and inside the star
can be estimated by
integrating in (\ref{eq:taudiff}) from the surface to the heater
and from the heater to the crust-core interface, respectively. The
outburst evolution
is determined by  $\tau_{\rm diff}^{\rm
out}$ and $\tau_{\rm diff}^{\rm in}$ as well as by the duration
$\Delta t$ of the internal outburst.
The time $\tau_{\rm diff}^{\rm out}$ regulates direct diffusion to
the surface that is most
important for $\Ls(t)$. Also, $\Ls(t)$ is indirectly affected by
$\tau_{\rm diff}^{\rm in}$ that mainly controls
the heat flow inside the star.

Note that calculated lightcurves
$T_{\rm s}(t)$ are accurately computed 
by our code at timescales longer than
the scale $t_{\rm diff,b}$ of heat diffusion through the heat blanketing
envelope ($\rho<\rho_{\rm b}$,
Section \ref{sec:level2}). This limits the time resolution of the code; 
$t_{\rm diff,b}$ depends on the choice of $\rho_{\rm b}$ (i.e., on geometrical
thickness of the heat blanket), on chemical composition of the
blanket and on the value of $T_{\rm s}$ (see, e.g.,
\citealt{2016BPY,2017Ch}); $\tau_{\rm diff,b}$ can be estimated from
the same equation~(\ref{eq:taudiff}).

For an $1.4$ \Msun\ star which has $R=10-12$ km and a non-redshifted surface
temperature $T_{\rm s}=10^6$ K, with the heat blanketing envelope
made of iron and extended to $\rho_{\rm b}=10^{10}$ $\gcc$,  we
obtain $\tau_{\rm diff,b}\sim 100$ d.
This time resolution of the code may be insufficient. 
Decreasing $\rho_{\rm b}$ to $10^9$ $\gcc$, we
obtain a better resolution, $\tau_{\rm diff,b} \sim 5$ d,
but calculations can be more time consuming.

In this section, we imply the heat blankets \citep{2016BPY}
made of iron and set $\rho_{\rm b}=10^{9}$ g cm$^{-3}$ 
for the $1.4\,M_\odot$ star, to reduce $t_{\rm diff,b}$, 
and $\rho_{\rm b}=10^{10}$ g cm$^{-3}$ for the $1.85\,
M_\odot$ star, that possesses thinner
crust with shorter $\tau_{\rm diff,b}$.

We set the heat power $Q(\rho, t)$ uniform inside the heater
and zero outside. Following \cite{2017Ch}, we vary $Q(\rho, t)$ as
\begin{equation}
Q(\rho, t) = H_{\rm 0} + (H_{\rm max}-H_{\rm 0})\sin ^{2} \left(
\frac{\pi t}{\Delta t}\right), \thinspace 0 \leq t \leq \Delta t,
\label{eq:QQQ}
\end{equation}
where $t$ is the {\it time from the beginning of the variation},
$\Delta t$ is the variation duration and $H_{\rm max}$ ($> H_0$) is the
maximum heat power reached at $t=\Delta t/2$. 
Then at $0 \leq t \leq \Delta t$ the total heat power
varies as
\begin{equation}
\Lh(t) = L_{\rm h0}^\infty + (L_{\rm h,max}^\infty-L_{\rm
h0}^\infty)\sin ^{2} \left( \frac{\pi t}{\Delta t}\right),
\label{eq:QQQQ}
\end{equation}
where $L_{\rm h,max}^\infty$ is the maximum integrated heat power
(\ref{eq:LLL}). When  $t$ exceeds $\Delta t$, we return the heater
to the stationary state, with $Q(\rho, T) = H_{0}$ and $\Lh=L_{\rm
h0}^\infty$. The star also begins returning to its
pre-outburst state.

It is also convenient to introduce the maxima of the surface
luminosities, $L_{\rm s,max}^{\rm \infty,SF}$ and $L_{\rm
s,max}^{\rm \infty,No SF}$, calculated with and without crustal
superfluidity, respectively, and the ratios
\begin{equation}
 \alpha_{\rm h}=\frac{L_{\rm h,max}^\infty}{L_{\rm h0}^\infty},
 \quad \alpha_{\rm s}=\frac{L_{\rm s,max}^{\rm \infty,SF}}{L_{\rm s,max}^{\rm
 \infty,No SF}}.
\label{eq:alpha}
\end{equation}
Here, $\alpha_{\rm h}>1$ is the ratio of the heat powers in the
outburst maximum and the steady state; $\alpha_{\rm s}$ is
the ratio of the maximum surface luminosities for a given superfluid
model (SF) and the non-superfluid crust (NoSF).

In this section, we set $H_{\rm 0} = 10^{17}$ erg cm$^{-3}$
s$^{-1}$, $\Delta t = 1$ yr, choose $\rho_1$ in the range from
$10^{11}$ $\gcc$ to $10^{13}$ $\gcc$, and choose $\rho_2$ in such a
way to have {\it one and the same} $\Lh$ for every heater's position
in a star with fixed $M$. For $\rho_1=10^{11}$ $\gcc$ we always set
$\rho_2=10^{12}$ $\gcc$. At $\rho_1=10^{12}$ $\gcc$ we have
$\rho_2=1.27 \times 10^{13}$ $\gcc$ ($L^\infty_{\rm h0} =2.77 \times
10^{34}$ erg s$^{-1}$) and $1.23 \times 10^{13}$ $\gcc$
($L^\infty_{\rm h0} =1.68 \times 10^{34}$ erg s$^{-1}$) for the
1.4 and 1.85~\Msun\ stars, respectively.

Our results are in line with those
in \cite{2017Ch}.
Fig. \ref{fig:main_cool} compares the outbursts of the
heat power $\Lh(t)$ and the surface luminosity $\Ls(t)$  (both in
units of $L_{\rm h0}^\infty$ with $\alpha_{\rm h}=500$) for the 1.4 and 1.85~\Msun\
stars. It displays  $\Lh(t)$ and $\Ls(t)$ for a non-superfluid crust
\citep{2017Ch} and for the three models of crustal superfluidity
(Fig.\ \ref{fig:Tcrit}). The left panel is for the
outer heater ($\rho_1 = 10^{11}$ $\gcc$);
the right panel is for the deeper one ($\rho_1 = 10^{12}$ g cm$^{-3}$).
The initial luminosity $L^\infty_{\rm s0}$ for the 1.85~\Msun\ star
is about two orders of magnitude lower than for the 1.4~\Msun\ star,
because the massive star undergoes much faster direct Urca neutrino cooling  
(Section \ref{sec:level2}). 

The propagation of heat from the heater to the surface
distorts the  $\Ls(t)$ profile with respect to $\Lh(t)$.
The $\Ls(t)$ profile is broader; its peak is reached
later. Our assumed $\Lh(t)$ profile is symmetric with respect to the peak
maximum, but the $\Ls(t)$ peak is asymmetric. The surface
outburst persists at $t>\Delta t$, when the internal
outburst is already over. 

First we focus on the
1.4~\Msun\ star. A rapid rise of $\Ls(t)$ after the onset of
the outburst is regulated by $\tau_{\rm diff}^{\rm out}$ in the
crust which is initially not very hot, so that the heat diffusion to
the surface is sufficiently fast. As a result of this rise-stage,
the heater and its vicinity become much hotter, which noticeably
blocks the heat diffusion. That is why the decay of $\Ls(t)$ after
the peak goes generally slower than the rise producing a strong
surface outburst asymmetry. 

If $\rho_1 =10^{11}$
$\gcc$ (the left panel of Fig.\ \ref{fig:main_cool}), the heater
penetrates into the inner crust but only slightly. 
An example of the deeper heater fully placed in the inner crust
is plotted on the right panel of Fig.~\ref{fig:main_cool} ($\rho_1=10^{12}$ $\gcc$). Since strong superfluidity reduces the heat
capacity of the inner crust, it increases the temperature in the
heater's vicinity after the internal outburst. Therefore, the
peaks of $\Ls(t)$ become relatively higher (it is easier to heat the crust),
and the post-outburst decay of $\Ls(t)$ relatively shorter
(lower $\tau_{\rm diff}^{\rm out}$), but in any case the rise time
is much shorter than the decay time. Actually, all
superfluidity effects are quite pronounced  for the deeper heater
but almost invisible for the outer heater. A discussion of similar
effects has been given by \citet{2009BC,2013PR}.

\begin{figure*}
\centering
\includegraphics[width=0.96\columnwidth]{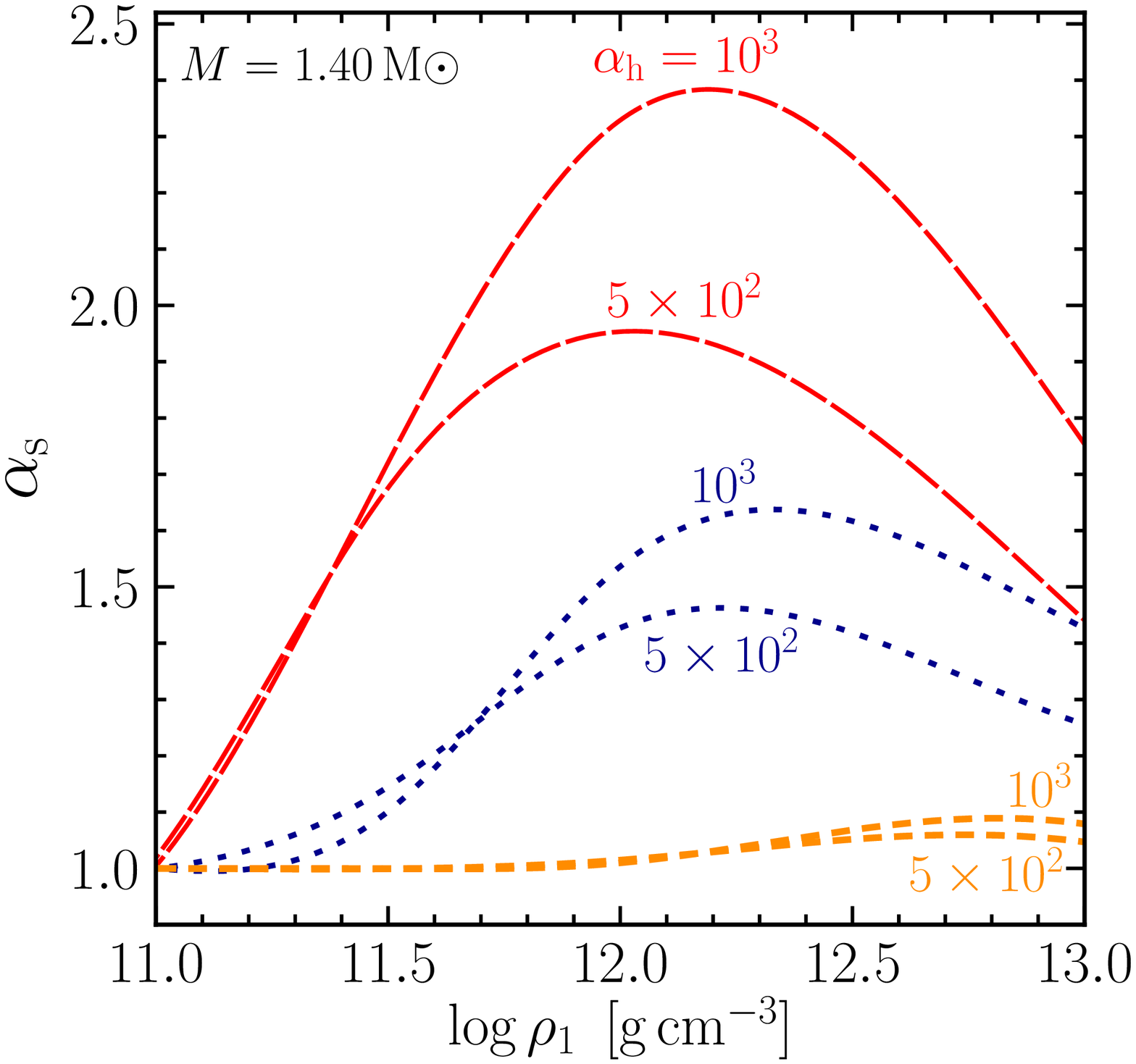}%
\includegraphics[width=0.96\columnwidth]{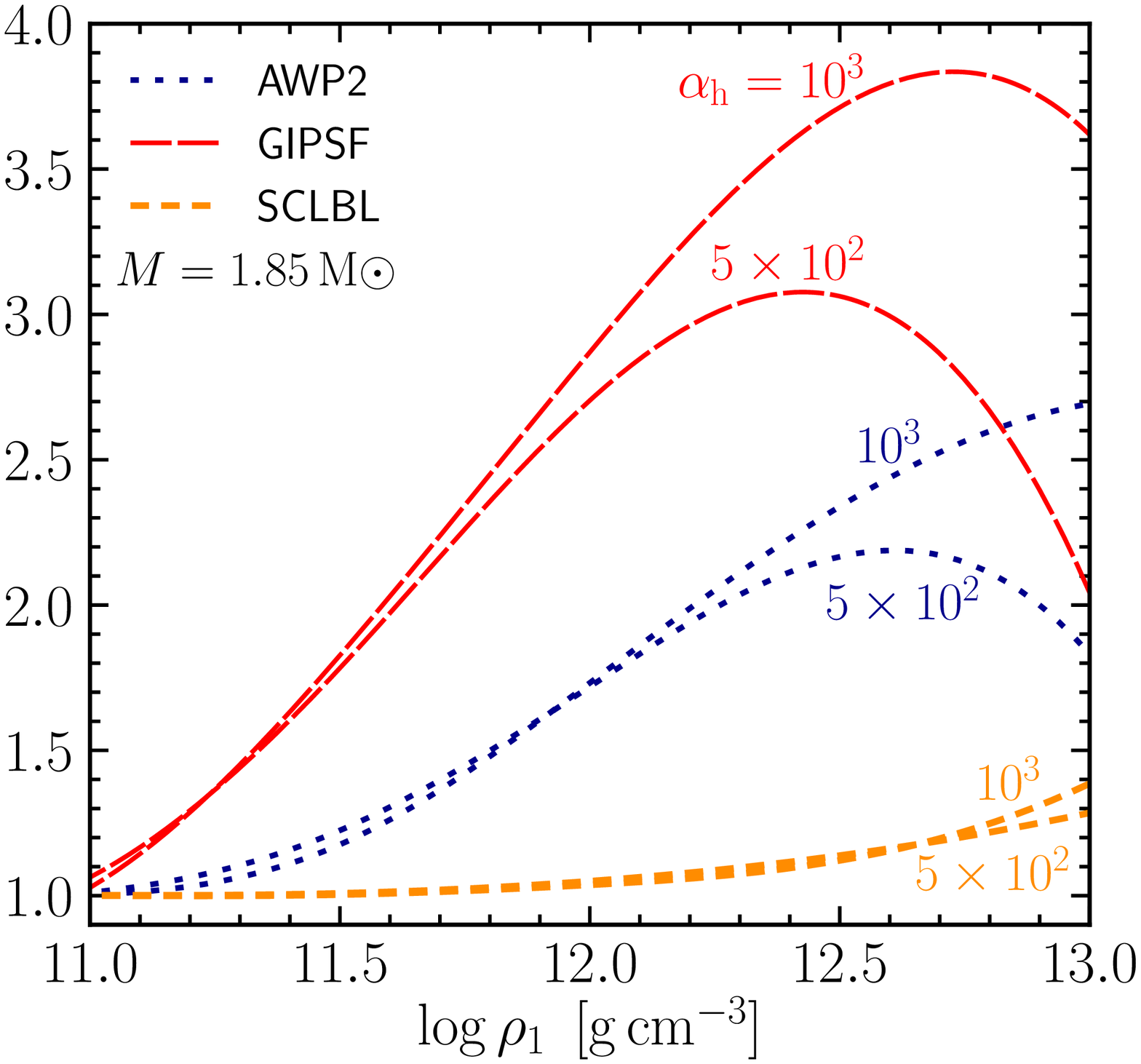}
\caption{\label{fig:general} Ratios $\alpha_{\rm s}$ defined by
equation (\ref{eq:alpha}) vs.\ $\log \rho_1$, for a lighter (left)
and heavier (right) star with the three superfluidity models at two
values of $\alpha_{\rm h}$. For fixed $\alpha_{\rm h}$, we obtain
three lines $\alpha_{\rm s}(\rho_1)$ (of different
types), depending on superfluidity model.}
\end{figure*}

The case of the 1.85~\Msun\ star in Fig.\ \ref{fig:main_cool} is
basically similar but the parameters are different. The star is more
compact than the  1.4~\Msun\ star and the crust is thinner which
reduces all diffusion time scales in the crust. In addition, recall that there
appears a powerful direct Urca neutrino cooling. 
All $\Ls(t)$ for the 1.85~\Msun\ star in Fig.\
\ref{fig:main_cool} correspond to the same $\Lh(t)$. The
intense neutrino cooling from the core of the 1.85~\Msun\ star lowers the core temperature
and creates much stronger temperature gradients between the heater
and the core than for the 1.4~\Msun\ star. This greatly
intensifies the heat outflow into the
core. Regardless of superfluidity, the diffusion time-scale
$\tau_{\rm diff}^{\rm out}$
becomes sufficiently short. The time dependence $\Ls(t)$ is
mainly regulated by heat diffusion through the outer crust.

In the left panel of Fig.\ \ref{fig:main_cool}, the $\Ls(t)$ peaks 
are more pronounced and less retarded from $\Lh(t)$ for
stars of both masses than in the right panel, since the heater
is geometrically closer to the surface.  As the
1.85~\Msun\ star has a thinner crust, its $\Ls(t)$ peaks are higher.
The difference between the luminosity peaks for the cases of
superfluid and non-superfluid crusts is significant only in the
right panel, where  $\rho_1 > \rho_{\rm drip}$ 
(e.g. \citealt{HPY2007}).  Accordingly, it is much easier to increase the
temperature inside and around the heater. Then more energy is left to heat
the surface. 

For a better understanding of Fig.\ \ref{fig:main_cool}, 
we plot additional Fig.\ \ref{fig:main_profile}
for the 1.4~\Msun. The figure shows the
internal temperature profiles $\Tg(\rho,t)$ throughout the crust. Here,
$\Tg=T \,\exp(\Phi)$ is the redshifted temperature $T(\rho,t)$ of
the matter. Note that $\Tg(\rho,t)=$ const in isothermal regions of the star at a
fixed $t$ (e.g., \citealt{1999REV} and references therein).
The profiles are
presented at four successive moments of time (0, 0.5, 3, and 15 yr), counted from the
beginning of the internal outburst.  
Now the initial steady-state heater ($t=0$) is not strong enough to destroy isothermality.
Other $T(\rho,t)$
profiles are bell-like, with maxima in the heater.  Similarly to Fig.\
\ref{fig:main_cool}, there is almost no difference (noticeable only
for the GIPSF model at $t \gtrsim 3$~yr) in the left panel
between superfluid and non-superfluid cases, because the heater is
placed mostly in the non-superfluid outer crust. Small changes in
the behavior of $\Tg(\rho,t)$ do occur, because the heat
fluxes within the inner crust are affected by superfluidity.
This effect is seen only after the internal outburst,
when the surface temperature starts to decline. For $\rho_1 =
10^{12}$ $\gcc > \rho_{\rm drip}$ in the right panel, the
difference in the temperature profiles of superfluid and
non-superfluid stars is quite visible. If the neutrons are
superfluid, the temperature rise is relatively fast and the $\Ls(t)$
peaks are higher. The highest $\Tg(\rho)$ peak and the most rapid
temperature growth occur for the GIPSF model.

Fig.\ \ref{fig:general} illustrates the effects
of different positions $\rho_1$ of the heater. 
It displays the ratios $\alpha_{\rm s}$, equation
(\ref{eq:alpha}), vs.\ $\rho_1$ for three models of superfluidity and
two values of $\alpha_{\rm h}$;  $\alpha_{\rm s}$
characterizes the enhancement of the $\Ls(t)$ peak by neutron
superfluidity. The ratios are larger for higher $\alpha_{\rm
h}$ because the contrast between the heat fluxes from the heater,
for superfluid and non-superfluid cases, increases as we enhance the
heat power. Another observation is that the $\alpha_{\rm s}$ maximum occurs at higher
$\rho_1$ as we increase $\alpha_{\rm h}$. If the heater is placed in
the outer crust, the effect of
suppressed heat capacity weakens, the difference of maximal
luminosities vanishes and all $\alpha_{\rm s}\to 1$. On the
contrary, if the heater is placed in the inner crust, more heat
disperses in the stellar interior and less heat reaches the surface,
decreasing $\alpha_{\rm s}$. The optimal case, with the maximum of
$\alpha_{\rm s}(\rho_1)$, occurs for intensive heaters in the inner
crust. Note that all optimal $\alpha_{\rm s}(\rho_1)$ are shifted to
higher $\rho_1$ for the 1.85~\Msun\ star because its crust is
thinner.

\begin{figure}[t]
\centering
\includegraphics[width=0.98\columnwidth]{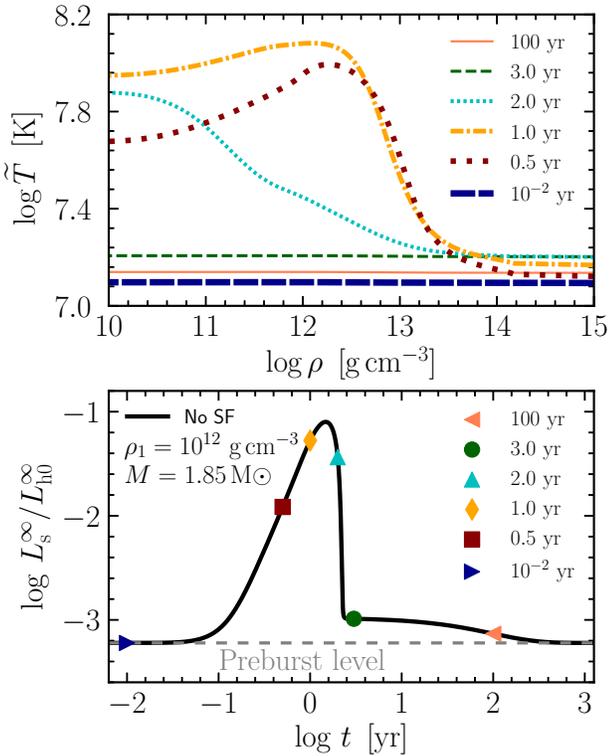} 
\caption{\label{fig:longest_term} Thermal evolution  of the
1.85~\Msun\ non-superfluid star with the inner heater
($\rho_1=10^{12}$ $\gcc$, $\Delta t=1$ yr) producing an outburst at
$\alpha_{\rm h}=500$; time $t$ is measured from the beginning of the
internal outburst. The upper panel shows $\Tg$ vs. $\rho$ at
different moments of time. The lower panel demonstrates the light
curve $\Ls(t)$ vs. $\log t$; the symbols indicate the moments of
time chosen on the upper panel (cf. the lower right panel in
Fig.\ \ref{fig:main_cool}). The dashed gray line would be
realized without any  outburst.}
\end{figure}

As seen from Fig.\ \ref{fig:main_cool}, the afterburst relaxation
of the 1.85 \Msun\ star has {\it finite duration} and stops
abruptly.  In Section \ref{sec:relax} we show that the same happens
for the 1.4 \Msun\ star but its relaxation lasts longer and
its final stage is not reproduced in Fig.\ \ref{fig:main_cool}.
The figure shows that the
initial luminosity $\Ls$ of the  1.85 \Msun\ star  at $t<0$ is lower than the luminosity $\Ls$
after the relaxation stops.  The internal thermal relaxation is achieved, 
but $\Ls$ (or, equivalently, $\Ts$) does not return to its
initial value. The star becomes hotter than before the outburst.
To explain this result, in Fig.\ \ref{fig:longest_term} we display
the dynamics of the outburst for the 1.85~\Msun\ fully
non-superfluid star with the inner heater ($\rho_1=10^{12}$ $\gcc$) at
$\alpha_{\rm h}=500$. The upper panel presents the density
dependence of the internal temperature $\Tg$ at different moments of
time, from $t=10^{-2}$ to $10^2$ yr. The lower panel shows the
observable lightcurve $\Ls(t)$. To guide the eye, the gray dashed
line shows $\Ls(t)$ without any outburst; it is the same as the
preburst level. The time scale is logarithmic and distorts 
the lightcurve profile but allows us to extend this profile to $t \approx
10^3$ yr. 

One can see the initial, nearly isothermal
interior just after the beginning of the outburst. It is violated by
the outburst in the crust for about 2.5 yr, longer than the outburst
duration $\Delta t=1$ yr, while the core stays isothermal and
becomes warmer by the heat flowing from the crust. In about 2.5 yr
(which is the time-scale of the {\it long-term relaxation} to be discussed
extensively in Sections \ref{sec:relax} and \ref{sec:level6}) the state of
isothermal interior is restored again, although the entire star gets warmer. 
This star is thermally relaxed inside and cools
slowly on much longer time scales of a few hundred years reaching
the preburst stage. It is a new epoch of passive cooling at which
the star `still remembers' the outburst heating.
This epoch mimics the stage of the second, {\it longest-term} relaxation.
Our additional calculations indicate that the new stage is
pronounced only in sufficiently cold and heavy stars with the open
direct Urca process in the core and with a powerful internal
outburst, capable to warm up the entire star.

\section{Deep crustal heating}
\label{sec:level6}

In this section, we present a schematic analysis of the crust-core
relaxation in neutron stars which enter LMXBs. A
neutron star in an LMXB transiently accretes matter from its
low-mass companion. There are active periods of
accretion with a lot of accretion energy released at the star surface. Then the LMXB is observed
as a bright X-ray source. After the accretion stops, the surface
energy release is off and a quiescent stage begins, with much weaker
surface thermal luminosity. Nevertheless, this quiescent emission is 
observable implying that the neutron stars remain warm and radiate
at a lower level until the next accretion episode.

It is widely thought that the 
quiescent thermal emission of LMXBs is mainly powered by the deep crustal
heating mechanism. The mechanism operates in a star's crust
during accretion periods, when the freshly accreted matter
compresses the underlying accreted material. The accreted matter
is known to be not in full equilibrium; it has a higher
energy than the fully equilibrated ground-state (cold-catalyzed) matter.
The accretional compression triggers nuclear
transformations which drive the matter to full equilibrium, with a modest total
energy release of 1--2 MeV per one accreted nucleon when the matter is compressed
to the densities $\sim 10^{13}$ $\gcc$. 
It is important, that this
energy is mainly generated at high compressions, deeply in the crust.
The energy generation was calculated by \citet{1990HZ} and refined
later by \citet{2008HZ,2007Gupta}. The deep crustal heating
mechanism was applied to LMXBs by \citet{1998BBR}. In principle, the
compression of the cold-catalyzed matter can also heat the star but
the effect is negligible.

The deep crustal heating creates a heater during  
accretion outbursts, with the energy generation rate
$Q(\rho,t)$ distributed over the accreted crust. In quiescence, the
compression due to the fall of newly accreted matter stops and 
$Q(\rho,t)=0$.
Initially, when an LMXB started to show
the accretion episodes, the star's crust was likely composed of
cold-catalyzed (ground-state) matter (e.g., \citealt{HPY2007}).
However, later, during many accretion episodes, it is gradually
replaced by a non-equilibrium accreted matter
\citep{1990HZ,1990HZa}, where the deep crustal heating can operate.
This replacement starts from the surface layers and moves toward the
crust-core interface (to $\rho_{\rm cc} \approx 1.5 \times 10^{14}$
$\gcc$). Nevertheless, in the deepest crust, $\rho \gtrsim 10^{13}$
$\gcc$, the difference between the accreted and cold-catalyzed
matter becomes so small that $Q(\rho,t)$ is almost negligible. After
the bottom of the accreted crust reaches the layer of the density
$\sim 10^{13}$ $\gcc$, the replacement can be treated as nearly
complete. A typical mass of the crust of the 1.4~\Msun\ star at
$\rho \lesssim 10^{13}$ $\gcc$ is about 0.05 per cent of the total
star's mass. For a typical {\it time-averaged} mass accretion rate
$\langle \dot{M} \rangle \sim 10^{-11} - 10^{-10}$ \Msun\
yr$^{-1}$ (e.g., \citealt{2015MB1}, and references therein) it
takes about $5 \times 10^6 - 5 \times 10^7$ yr to replace the
cold-catalyzed crust by the accreted one. Before that, the accreted
crust can be thinner and the deep crustal heating weaker. In
principle, one should consider the deep crustal heating with fully
accreted or partly accreted (hybrid) crusts. The possibility of
hybrid crusts has been studied by
\citet{2013SXRT} to explain the anomalous observational behavior of
the IGR J17480--2446 source.

\begin{figure}[!t]
\centering
\includegraphics[width=0.98\columnwidth]{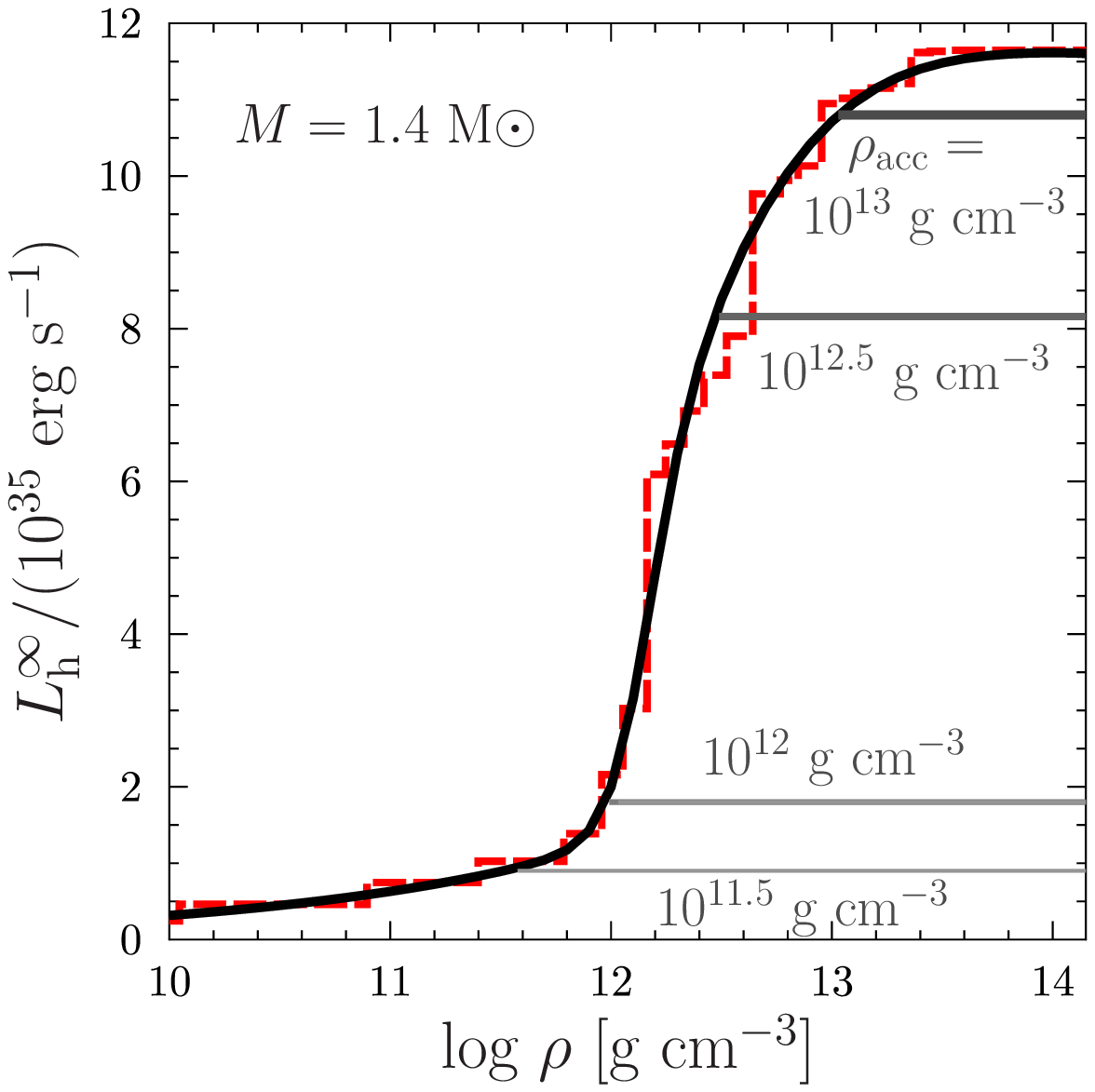}
\caption{\label{fig:Qrho} Generated heat power $\Lh(\rho)$,
integrated over a spherical layer from the surface to a 
density $\rho$, in the crust of the 1.4 \Msun\ star
at $\dot{M}=1.5 \times 10^{-8}$ \Msun\ yr$^{-1}$. The broken dashed line corresponds
to the model of deep crustal heating by \citet{2008HZ} for the fully
accreted crust. The thick black curve is the smoothed curve used in
the calculations. Four horizontal gray curves of different
thicknesses show the modifications of the smoothed $\Lh(\rho)$ curve
at $\rho>\rhoacc$, with
$\rhoacc=10^{11.5}$, $10^{12}$, $10^{12.5}$ and
$10^{13}$ $\gcc$. A rapid growth at $\rho \sim 5 \times 10^{11}$ $\gcc$ is
due to the neutron drip. }
\end{figure}

For illustration, Fig.\ \ref{fig:Qrho} demonstrates the heat
power $\Lh(\rho)$, given by equation (\ref{eq:LLL}) and integrated
from the surface to a layer with a given $\rho$ in the crust, for
a fixed  $\dot{M}=1.5 \times 10^{-8}$~\Msun\ yr$^{-1}$ during an accretion
stage. Since $\Lh(\rho)$ is
directly proportional to $\dot{M}$, one can easily rescale
$\Lh(\rho)$ to any $\dot{M}$. We take the same 1.4~\Msun\
star's model, as before. The dashed line corresponds to the
deep crustal heating for the fully accreted crust, with $\rhoacc \approx \rho_{\rm cc}$, 
and the mass of the accreted crust $\Delta M_{\rm acc} \approx
0.02$ \Msun. The heat distribution $Q(\rho)$  within the crust 
is taken from \cite{2008HZ}. 
There are many thin shells (steps of the dashed
line) where the heat is released; the actual function $Q(\rho)$
looks as a sequence of narrow peaks. For simplicity, we replace the
stair-step-like heat power $\Lh(\rho)$ by a smooth curve with nearly
the same integrated $\Lh(\rho)$ for the fully accreted crust; it is
shown by the thick black curve.

However, if the accreted crust extends to lower $\rhoacc< \rho_{\rm
cc}$, our basic smoothed black curve $\Lh(\rho)$ is modified -- it
becomes constant, $\Lh(\rho)=\Lh(\rhoacc)$, at $\rho
> \rhoacc$. These modified parts of the
$\Lh(\rho)$ curves are shown by the gray horizontal lines of
different thickness for the four bottom densities,
$\rhoacc=10^{11.5}$, $10^{12}$, $10^{12.5}$ and $10^{13}$ g
cm$^{-3}$ of the accreted crust (with the four masses
$\Delta M_{\rm acc}= 3.3 \times 10^{-5}$, $6.2 \times 10^{-5}$, $1.7
\times 10^{-4}$ and $7.3 \times 10^{-4}$ \Msun). If
$\rhoacc>10^{13}$ $\gcc$, the crust behaves as almost fully accreted. 
If $\rhoacc$ were about 3 times lower
($\rhoacc=10^{12.5}$ $\gcc$), the total heat generation would be
1.4 times lower than for the fully accreted crust. A further
reduction to $\rhoacc=10^{12}$ $\gcc$ or to $\rhoacc=10^{11.5}$
g~cm$^{-3}$ $\approx \rho_{\rm drip}$ would reduce $\Lh$,
respectively, by a factor of 6.5 or 12, as compared to the fully
accreted crust. However, in order to form such a thin accreted crust
one needs much less time than to form the fully accreted crust; the
chances to observe a star with a thin accreted crust seem small.

All LMXBs whose active and quiescent emission is observed can
be divided into two groups. The summary of observations has
been recently reviewed by \citet{2017SRTSobs}.

Group 1 contains the sources where outbursts are relatively short
(days--weeks) and/or not intense, with the mass
accretion rates $\dot{M}$ during the outbursts well below the
Eddington limit. Then the deep crustal heating is not strong enough
to destroy the isothermality of stellar interiors supported by
a high thermal conductivity. These sources can be
called {\it quasi-stationary} (e.g., \citealt{2017HS}).  
When the star transits from an active
state to quiescence, the observed surface luminosity drops quickly
to the quiescent level. In quiescence, the star 
undergoes passive cooling on
timescales about $10^3-10^4$ yr until the next outburst occurs in
months or years.  The
surface is thermally coupled to the stellar core. At this stage, a
quasi-stationary LMXB serves as a natural laboratory to probe the
physics of neutron star cores (e.g., \citealt{BULK}).

Group 2 consists of the sources demonstrating sufficiently long
(months--years) and intense (at about the Eddington level)
outbursts. During an outburst, the crustal heater is strong and
makes the crust hotter than the core, destroying the isothermality
of stellar interiors; such LMXBs were dubbed {\it
quasi-persistent} (e.g., \citealt{2012PR,2016Merritt}).
There appear large temperature gradients in the
crust near the boundaries of the heated layer. When the accretion
stops and the star transits to a quiescent state, the transition
takes more time (typically, a few years) than in quasi-stationary
LMXBs, because it includes an additional crust-core relaxation of
the overheated crust with the colder core (e.g.,
\citealt{2007Shternin,2009BC}). At such transition stages, the
stellar surface is thermally decoupled from the core but coupled to
the heater, so that neutron stars serve as natural laboratories to
study the physics of the crust. These transitions have much in
common with the initial crust-core relaxation of newly born isolated
neutron stars (e.g., \citealt{1994RELAX,Gned2001}). After the end of such a
{\it long-term} crust-core relaxation in a transiently
accreting star, the star reaches a quiescent state with isothermal interior
and then resembles quiescent neutron stars in quasi-stationary
LMXBs; the physics of the crust becomes almost unimportant.

Note that our description of accretion and quiescent states is
idealized. In accretion stages, one observes the variability of $\Ls(t)$
produced by fluctuations of the accretion rate $\dot{M}(t)$ and by
thermonuclear outbursts in the subsurface layers. During quiescent
stages, $\Ls(t)$ can also vary because of sporadic low-level
accretion and thermonuclear outbursts (e.g. \citealt{2014HOMAN} and
references therein).

\begin{figure*}[!t]
\centering
\includegraphics[width=0.46\textwidth] 
{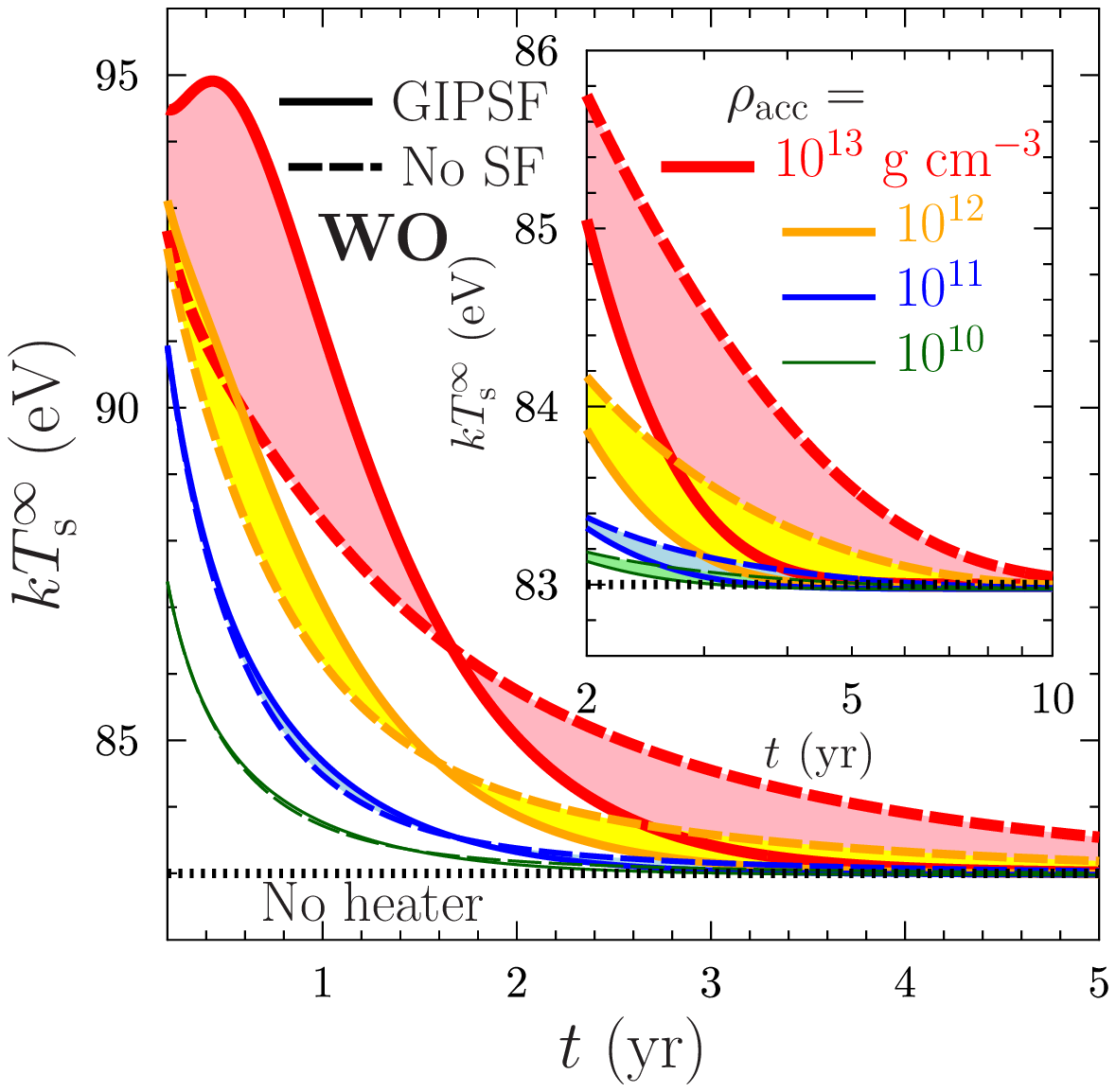}%
\hspace{5mm}
\includegraphics[width=0.46\textwidth] 
{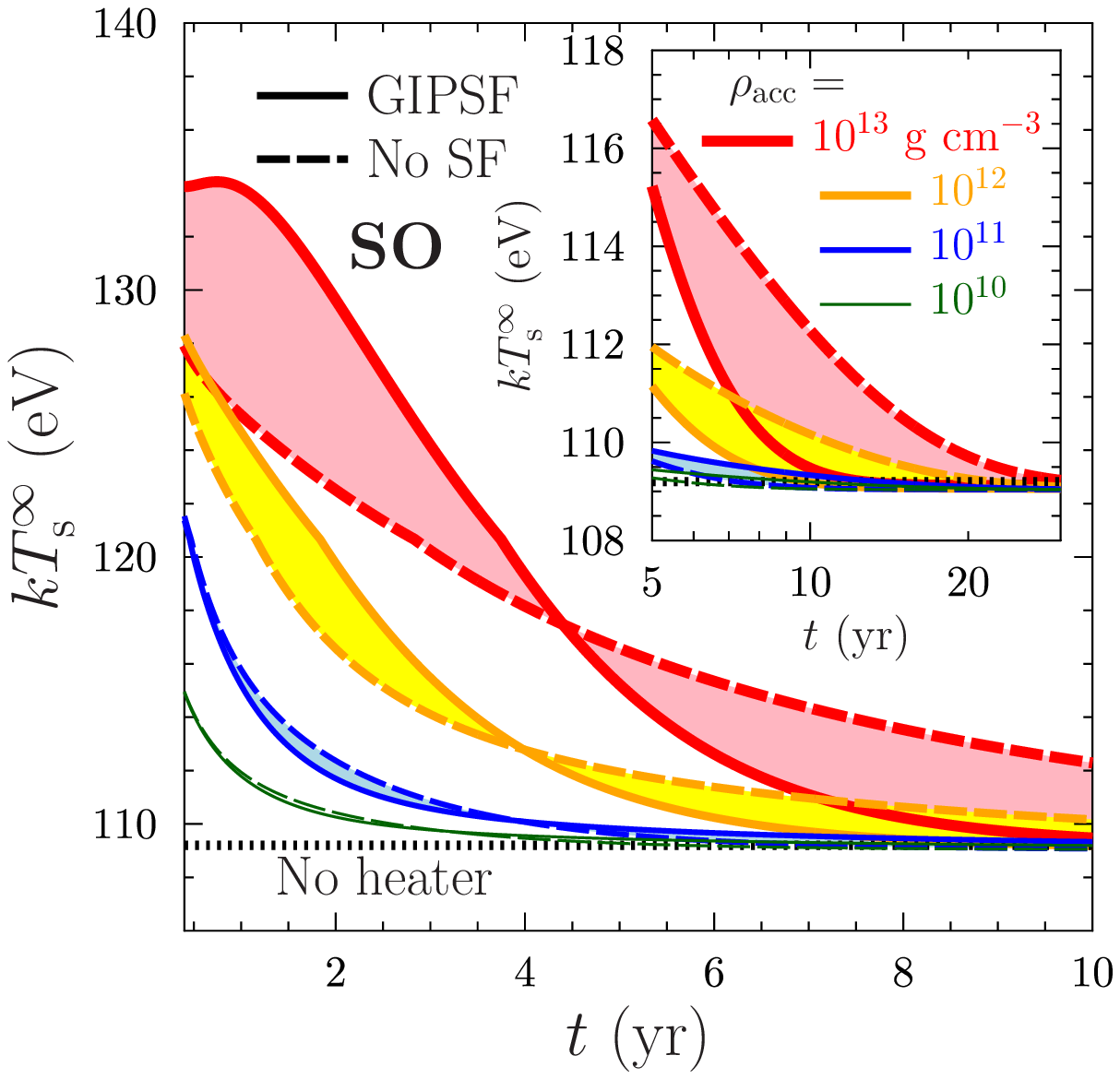}%
\caption{\label{fig:relax} The dependence of $\Ts$ for the
1.4\, \Msun\ neutron star on time $t$ since the
outburst for the models of weak (left) or strong (right)
outbursts (WO or SO, Table \ref{tab:1}).
The four solid curves of different thicknesses on
each panel correspond
to $\rho_{\rm acc}=10^{10},~10^{11},~10^{12}$ and $10^{13}$ $\gcc$ 
(cf. Fig.\ \ref{fig:Qrho}) and the
GIPSF model for crustal superfluidity. The four dashed curves refer
to the same models but without 
superfluidity. Shaded areas can be filled by $\Ts(t)$ curves for  superfluidities of intermediate strength. 
The surface temperature without any heater is almost
constant (the dotted horizontal lines). The insets visualize the
very late stages of the long-term relaxation. 
}
\end{figure*}

Here we focus on the {\it long-term relaxation} from accretion to
quiescence in the quasi-persistent LMXBs. They have been observed 
from several LMXBs (e.g., \citealt{2014HOMAN,2017SRTSobs}).
There have been many attempts to
model long-term cooling in the quasi-persistent sources
(see, e.g.,
\citealt{2007Shternin,2009BC,2012PR,2013PR,2015TAP,2014MC,2014HOMAN}).
The main conclusion is that the deep crustal heating reasonably well
describes the late stages  of the long-term relaxation -- a few
months--years after an outburst.

However, the deep crustal heating is usually insufficient to explain
the early transition stage -- a few first hundred days in
quiescence. To reproduce such a stage, one assumes an
additional heat source in the outer layer of the crust (as was first
suggested by \citealt{2009BC}). The nature of this special 
thermal evolution of the outer layers is not clear. It is often addressed 
as {\it shallow heating} and treated phenomenologically.

\section{Long-term relaxation in quasi-persistent LMXBs}
\label{sec:relax}

Let us describe our simulations of long-term cooling of neutron stars in
quasi-persistent LMXBs. We use the same cooling
code as described above for the same 1.4\, \Msun\ 
star model. The heater's power is now
distributed over a thick crustal layer (Fig.\ \ref{fig:Qrho}). Our
calculations may be inaccurate at the beginning of the quiescent
stage because we neglect the shallow heating. 
However, they are expected
to be adequate to reproduce the later long-term 
relaxation in quiescence. The
surface temperature at this stage is mostly
regulated by the heating in the deep crust, being 
rather insensitive
to the shallow heating (e.g. \citealt{2009BC}).

\renewcommand{\baselinestretch}{1.2}
\begin{table}
	\caption{Two outburst models for 1.4 \Msun\ neutron star} \label{tab:1}
	\begin{tabular}{llll}
		\hline \hline
		Model  of &  $\Tg_0$ & $\dot{M}^\dagger$ & $\Delta t$\\
		outburst       & K        & \Msun\ yr$^{-1}$ & yr  \\
		\hline
		Weak (WO) & $5.7 \times 10^{7}$ & $8.4 \times 10^{-9}$ & 0.2  \\
		Strong (SO) & $1 \times 10^{8}$ & $1.5 \times 10^{-8}$ & 1.0  \\
		\hline \hline
	\end{tabular} \\
	{\small $^\dagger \dot{M}$ 
	is the mass accretion rate during an outburst}
\end{table}
\renewcommand{\baselinestretch}{1.0}

Note that one should not confuse the models of accreted
heat blanketing envelopes ($\rho < \rho_{\rm b}\lesssim 10^{10}$
$\gcc$) and accreted crust ($\rho < \rhoacc$). A heat blanket
is thin and situated just under the stellar surface. Its
composition is mostly regulated by thermonuclear burning of freshly
accreted material during outbursts (e.g., \citealt{2002Brownetal,2006SB}). 
An accreted crust is thicker and extends much deeper than the heat
blanket. Its composition is determined by
temperature-independent nuclear transformations in the deep layers
(e.g., \citealt{1990HZ,2008HZ}). In quiescence, the heat blanket
may contain iron but the crust beneath it can have
accreted composition \citep{1990HZ}.

The simulations have been performed in a slightly
different manner than in Section \ref{sec:level4}.
First, an initially hot star is cooled freely, with
$\Lh=0$. After the internal thermal relaxation is over and the
temperature $\Tg$ reaches some predetermined value $\Tg_0$, 
we switch on the heater in the
accreted crust (Fig.\ \ref{fig:Qrho}) assuming a certain mass accretion
rate $\dot{M}$ during an outburst
stage. The surface luminosity $\Ls(t)$ calculated by the code
during this outburst is smaller than the actual luminosity 
because the calculated $\Ls(t)$ does not include
a huge surface energy release due to accretion. Nevertheless,
the internal temperature $\widetilde{T}(\rho,t)$ in the deep crust
and in the core as well as the long-term $\Ts(t)$ relaxation
can be calculated in this way.

Let us stress that the problem in Section~\ref{sec:level4} was 
more ``academic.'' We studied an inner heater of unknown nature
as the {\it only} heat source. Here we use the well elaborated deep
crustal heating mechanism. It {\it should be accompanied by
the accretion heating from the surface, and most probably by the
shallow heating} but we do not include these two additional mechanisms.

\begin{figure}[!t]
	\centering
	\includegraphics[width=0.46\textwidth]{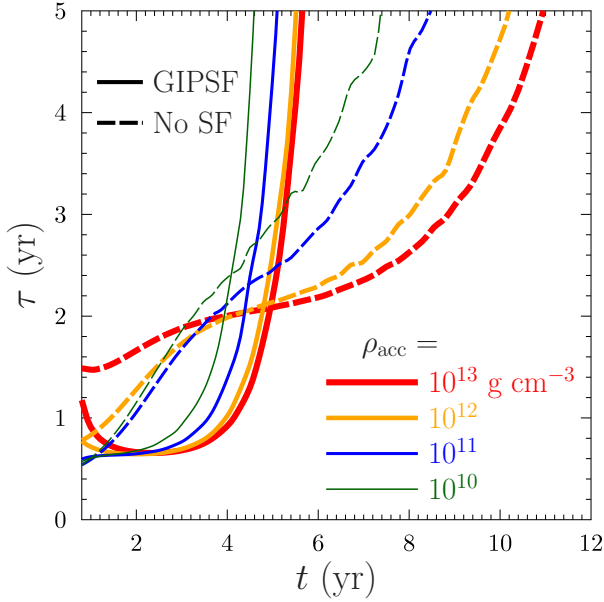}
	\caption{\label{fig:relaxtaun} Time dependence of the local 
		relaxation time $\tau(t)$ 
		for the WO case in left panel of Fig.\ \ref{fig:relax};
		see text for details.
	}
\end{figure}

At a certain moment of time $\Delta t$ after the outburst onset,
we abruptly switch off the deep crustal heating to simulate the
relaxation. We
have $L_{\rm h}^\infty=0$ in quiescence and a constant $\Lh$ during
the outburst. Any $\Ls(t)$ lightcurve is then specified by
$\widetilde{T}_0$, by the heat power
$\Lh$ and $\Delta t$. Of course, our
lightcurves depend also on assumed models of the star, heat blanket
and crustal superfluidity. 

\begin{figure*}[!t]
	\centering
	\includegraphics[width=0.46\textwidth] 
	{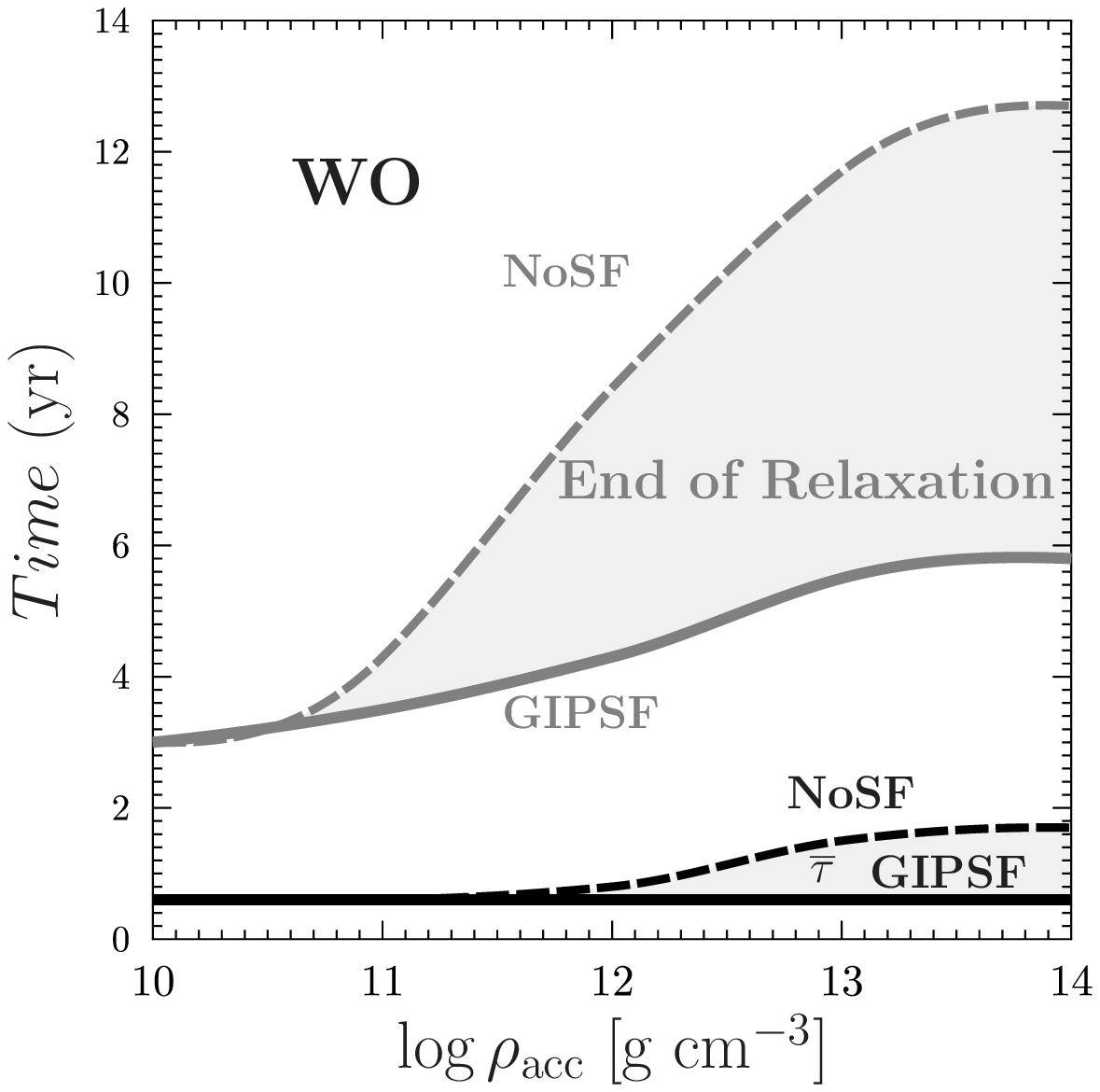}%
	\includegraphics[width=0.46\textwidth] 
	{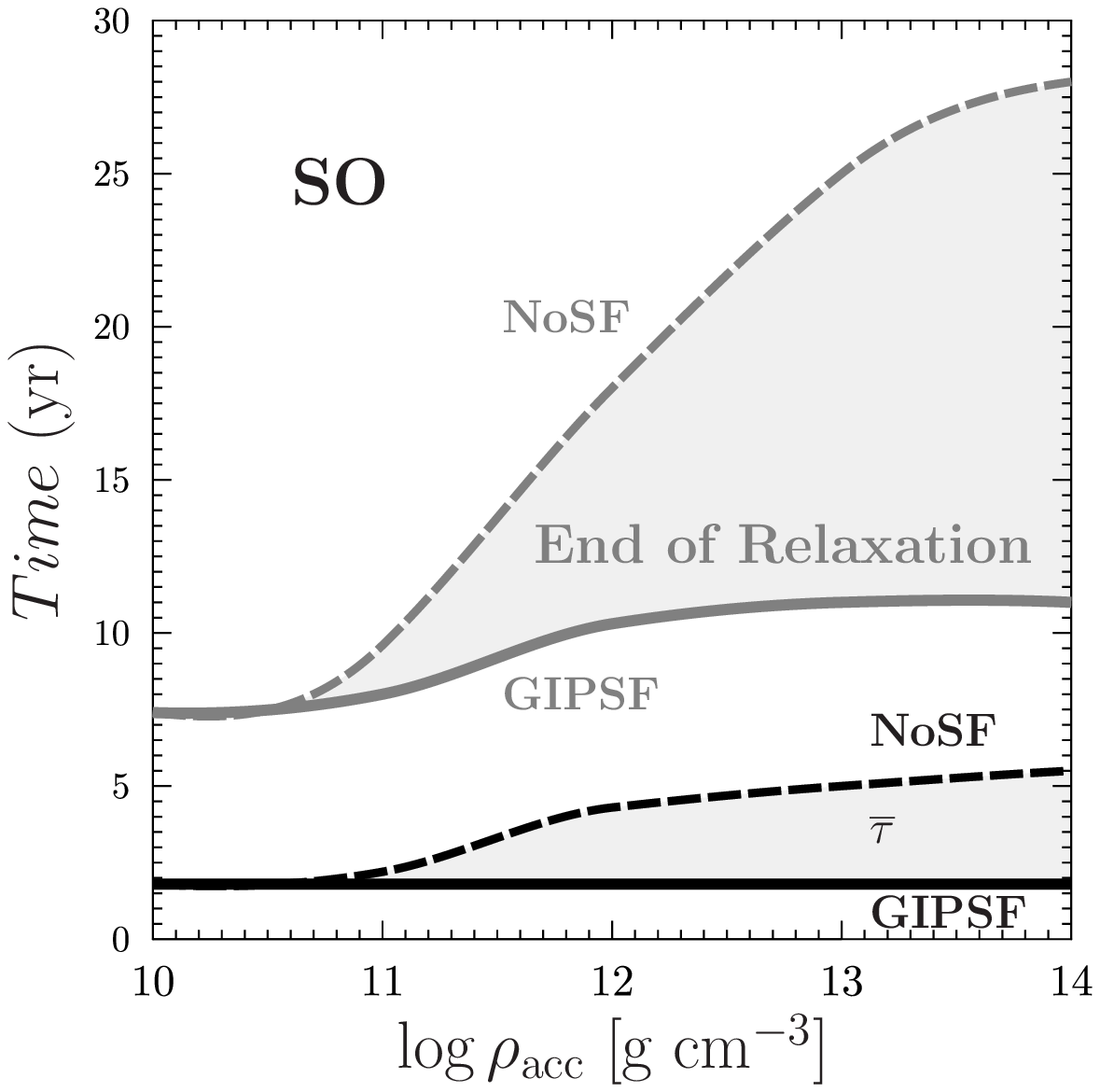}%
	\caption{\label{fig:relaxun} Characteristic relaxation
	times $\overline{\tau}$ (black curves) and relaxation durations
	$t_{\rm rel}$ (grey curves) versus $\rhoacc$ for the WO (left) and SO 
	(right) models.  The solid and dashed lines are for the GIPSF superfluidity and non-superfluid crust, respectively. See text for
	details.
	}
\end{figure*}

For illustration, we took, rather arbitrarily, 
two models of the outburst, the weak outburst
(WO) and the strong outburst (SO). The parameters of these outbursts
are listed in Table \ref{tab:1}. The SO lasts 5 times longer and
has higher $\dot{M}$ (close to the Eddington limit) than the WO. 
The total deep crustal energy input
during the SO is 8.8 times larger. For each outburst
model we consider the cases of the non-superfluid and highly
superfluid (GIPSF) crust. The cases of weaker
superfluidities are intermediate between these two. 
We calculate the lightcurves $\Ts(t)$ for four
densities at the bottom of the accreted crust,
$\rhoacc=10^{10},\,10^{11},\,10^{12}$ and $10^{13}$ $\gcc$.  
If $\rhoacc=10^{10}$ or $10^{11}$ $\gcc$, the heater is fully placed
in the outer crust. Otherwise, it extends to the inner crust
that can be superfluid. Here we use the model of fully
accreted heat blanketing envelope of \citet{1997POT}. 

The calculated lightcurves $\Ts(t)$ (with $t$ being time since
the end of accretion) are plotted in Fig.\ \ref{fig:relax}; $\Ts$ is expressed in energy units, $k \Ts$, 
$k$ being the Boltzmann constant. 
The left and right panels refer to the
WO and SO, respectively.
The insets visualize the final stages of the long-term relaxation.
The solid and dashed curves correspond to the GIPSF 
and non-superfluid scenarios, 
respectively.

With increasing time $t$ after an outburst, the star relaxes
to the quasi-stationary state that would take place without heating 
(with isothermal interior at a certain
temperature $\Tg_1$). Since we do not consider a massive
star, we should obtain the solution with 
$\widetilde{T}_1 \approx \widetilde{T}_0$ 
(Section \ref{sec:level4}). We assume here that the characteristic
time-scale of passive cooling ($\gtrsim 1$ kyr) is much longer
than the time interval between two successive outbursts.
For the assumed heat blanketing model after the outburst 
we have $kT_{\rm s1}^\infty$ = 83 eV ($\Tg_0=57$ MK, Table \ref{tab:1})
for the WO, and $kT_{\rm s1}^\infty$ = 109 eV ($\Tg_0=100$ MK)
for the SO. 

According to Fig.\ \ref{fig:relax}, a thinner accreted crust 
reduces $\Ts$.  In any
case, the relaxation does not last longer than a few decades.  
In the absence of the heater (the dotted
horizontal line) the passive cooling would be unnoticeable
on the relaxation timescales. 

For a star with a massive superfluid accreted crust, 
the crust is especially
hot and the $\Ts(t)$ curves become non-monotonic reaching a peak at
$t \lesssim (1-2)$ yr. This non-monotonic behavior
mainly contradicts observations. It appears because we neglect the shallow heating
in our modeling. 

However, it is expected
(e.g., \citealt{2009BC}) that the shallow heating 
manifests itself just after the outburst (when the heat
emitted from the surface emerges from the shallow depths) and 
does not affect the long-term relaxation of our interest
(when the emitted heat emerges from the deeper crust).
The exclusion concerns the cases of $\rhoacc \lesssim 10^{10}$ $\gcc$,
in which the layers of deep crustal heating and the 
shallow heating essentially overlap. We include these cases only for
completeness of the consideration.

For the conditions in Fig.\ \ref{fig:relax} in the massive superfluid accreted
crusts, the long-term relaxation after the WO starts 
at $t \gtrsim 1$ yr. For a less massive
and/or non-superfluid crust it starts earlier. For larger $\Delta t$
and/or $\dot{M}$ (larger total outburst energy) it starts later.
For instance,  for the SO it starts at $t \gtrsim 1.6$ yr.

As seen from Fig.\ \ref{fig:relax}, there are two  
long-term relaxation regimes. As long as the accreted crust does not extend
beyond the neutron drip density ($\rhoacc < \rho_{\rm drip}$; that is for
$\rhoacc=10^{10}$ and $10^{11}$ $\gcc$ in our case), 
the lightcurves are affected
by superfluidity rather weakly. This is natural because in such cases the generated heat flows
to the surface through the non-superfluid outer crust. In the opposite case
($\rhoacc > \rho_{\rm drip}$, that is for
$\rhoacc=10^{12}$ and $10^{13}$ $\gcc$) the
heat is mainly generated in the superfluid region, 
the lightcurves are strongly affected
by superfluidity which greatly accelerates the relaxation.  

Let us stress that the existence of superfluidity in 
neutron star crusts in quasi-persistent LMXBs has been studied
previously, mostly, for fully accreted crusts (e.g.,
\citealt{2007Shternin,2009BC}). Our results agree with those studies.


Using the theoretical lightcurves on Fig.\
\ref{fig:relax}, we can analyse temporal behavior of 
$\Ts(t)$ during the long-term relaxation in quiescence.
It is instructive to write
\begin{equation}
    \Ts(t)=T_{\rm s1}^\infty+ \delta \Ts(t),
\label{eq:relax}		
\end{equation}
where $\delta \Ts(t)$ is a variable component of $\Ts$, whose
behavior is most important. One often describes 
this behavior (e.g., \citealt{2014HOMAN}) as exponential, 
$\delta \Ts(t) \propto \exp(-t/\tau)$,
or as power-law, $\delta \Ts(t) \propto t^{-n}$, with constant relaxation time
$\tau$ or (broken) power-law index $n$. 

To check the exponential behavior 
with our theoretical light\-curves, we have calculated the real local 
time-dependent $\tau(t)$ as 
\begin{equation}
  \tau(t)=-\frac{\delta \Ts(t)}{\delta \dot{T}^{\infty}_{\rm s} (t)},
	\label{eq:taun}
\end{equation}
for the WO lightcurves 
on the left panel of Fig.~\ref{fig:relax}. 

The results are presented in
Fig.\ \ref{fig:relaxtaun}. Evidently, the 
approximation of constant $\tau$ is inaccurate.  

Moreover, the $\tau(t)$ curves do depend on crustal
superfluidity. This is natural for $\rhoacc \gtrsim \rho_{\rm drip}$
but less evident 
for $\rhoacc \lesssim \rho_{\rm drip}$
because in the latter case the long-term lightcurves are weakly
affected by superfluidity (Fig.\ \ref{fig:relax}). Nevertheless,
this is true even in the latter case. 
This means that $\tau(t)$ curves (determined by
the time derivative $\dot{T}_{\rm s}^\infty(t)$)
are basically much
more sensitive to crustal superfluidity than the lightcurves themselves.
This is likely because of complicated character of heat diffusion
which ``feels'' the presence of superfluidity even if the superfluid layer
is placed deeper than the heat source. Therefore, 
$\dot{T}_{\rm s}^\infty(t)$ contains important
information on the physics of the crust, although extracting
this information from observations would be a
problem.     

Note also wave-like features of the curves in Fig.\
\ref{fig:relaxtaun}.  They are especially pronounced for nonsuperfluid
crust with $\rho_{\rm acc} \lesssim \rho_{\rm drip}$.
They are possibly associated with low-amplitude thermal waves which
travel within the crust in the vicinity of the heater, being supported
by weak reflections from the neutron drip zone and from the artificial
bottom of the heat blanket ($\rho=\rho_{\rm b}$).
This is another indication that the $\dot{T}_{\rm s}^\infty(t)$ curves
carry potentially important information on the crust structure.

The properties of $\tau(t)$ curves in
Fig.\ \ref{fig:relaxtaun} are as follows. There are wide minima
of $\tau(t)$ at $ t \lesssim 2.5$ yr for superfluid crust
with any $\rhoacc$.
As long as $t \gtrsim 2.5$ yr, all curves increase with time. 
Very roughly, any $\Ts(t)$ curve can be specified by 
a typical relaxation time $\overline{\tau}$
and a typical relaxation duration $t_{\rm rel}>\overline{\tau}$.
The relaxation time $\overline{\tau}$ corresponds to the minimum 
of $\tau(t)$ in Fig.~\ref{fig:relaxtaun}. Typical time 
$t=t_{\rm rel}$ of the later rapid
$\tau(t)$ increase marks
the approach of $\Ts(t)$ to $T_{\rm s1}$. It
confirms that the relaxation duration $t_{\rm rel}$ is finite. 
It is seen also in 
previous computer simulations of the
lightcurves $\Ts(t)$; e.g. 
\citet{2007Shternin,2009BC,2013PR,2014HOMAN,2016Merritt}. 
We have already discussed finite $t_{\rm rel}$ in Section
\ref{sec:level4} for the 1.85 \Msun\  star where the effect
is quite pronounced. 
Evidently, similar effect occurs also in
the 1.4\, \Msun\ star, although it is less pronounced. 

 A schematic
dependence of $\overline{\tau}$ and $t_{\rm rel}$ 
on $\rhoacc$ is shown in Fig.\ \ref{fig:relaxun} for the same conditions
as in Fig.\ \ref{fig:relax}. The left and right panels refer,
respectively, to the WO and SO models. The black lines show
$\overline{\tau}(\rhoacc)$ while the grey lines exhibit $t_{\rm rel}(\rhoacc)$.
The solid curves are for the GIPSF superfluidity, while
the dashed curves are for the non-superfluid crust. The shaded
regions, again, can be filled by respective curves for intermediate
superfluidities. 

We estimated $\overline{\tau}$ 
as the minimum $\tau(t)$ (fastest decay rate) at
the long-term relaxation stage (at
$t>1$ yr for the WO and $t>2$ yr for the SO). 
We have also calculated $\overline{\tau}$ in another way,
by taking theoretical lightcurves $\Ts(t)$ and fitting
them by the function $\Ts(t)=B+A\,\exp(-t/\overline{\tau})$ within a reasonable
time interval ($t_1 \leq t \leq t_2$). 
We have varied $t_1$ and $t_2$ within some limits. The choice 
of these limits does affect the inferred values of $\overline{\tau}$
but the dependences $\overline{\tau}(\rhoacc)$ stay qualitatively
the same as in Fig.\ \ref{fig:relaxun}.

The relaxation durations
$t_{\rm rel}$ in Fig.\ \ref{fig:relaxun} 
are estimated by eye as times
at which $\Ts(t)$ approaches $T_{\rm s1}^\infty$.

The relaxation time
$\overline{\tau}(\rhoacc)$ for the WO in the superfluid crust 
(the solid black curve
in the left panel of Fig.\ \ref{fig:relaxun}) 
with all $\rhoacc$ are about the same, $\sim  0.6$ yr. 
The presence of superfluidity  speeds up 
the heat diffusion through the inner (superfluid) crust, 
so that the relaxation time is mostly determined by the 
heat diffusion in the outer (non-superfluid) crust.
Accordingly, the diffusion time from the heater to the surface
depends on $\rhoacc$ but only slightly. Decreasing $\rhoacc$ from
$10^{13}$ to $10^{10}$ $\gcc$ lowers $t_{\rm rel}$ from
$\sim 6$ to $\sim 3$ yr, because the heater
becomes effectively closer to the surface. The relaxation time
$\overline{\tau}(\rhoacc)$ for the superfluid crust and the SO in Fig.\ \ref{fig:relaxun}
behaves similarly to the WO curve but is somewhat longer.

For the non-superfluid crust, the situation is different.
At $\rhoacc \lesssim \rho_{\rm drip}$, 
the relaxation time $\overline{\tau}(t)$ for the WO and SO stays about the same as in
the superfluid case. In contrast, for $\rhoacc \gtrsim \rho_{\rm drip}$, 
both times -- $\overline{\tau}$
and $t_{\rm rel}$ -- become a few times longer than in the superfluid
crust because the absence of superfluidity delays propagation of heat
to the surface. 

The finite duration $t_{\rm rel}$ of the long-term relaxation can be seen in other numerical
simulations of lightcurves in quiescent stages of quasi-persistent
LMXBs (e.g., \citealt{2009BC}). Superfluidity can shorten
$t_{\rm rel}$ by a factor of 2--3.

We have not attempted to tune the WO and SO parameters to explain specific quasi-persistent
LMXBs. Nevertherless according to the data (e.g., \citealt{2014HOMAN},
 particularly,
their Fig.\ 5; \citealt{2017SRTSobs}), the WO model can be successful to explain
the IGR J17480--2446 source while the SO model seems relevant for explaining XTE J1701--462.
For example, our very preliminary analysis indicates that if we apply the WO model for
the interpretation of the observations of the IGR J17480--2446 after
the 2010 October outburst, it is
profitable to assume that the crust is hybrid 
(partly accreted), with $\rhoacc \sim 10^{12}$ $\gcc$, and the inner crust
is highly superfluid. The possibility that the crust of the neutron star
in this object is hybrid has been discussed in the literature
\citep{2012Patrunoetal,2013SXRT}. More work is required to apply our results for a detailed
interpretation of various sources.

\section{Conclusions}
\label{sec:level5}

We have modeled thermal evolution of neutron stars with internal
spherically symmetric heaters located in the  crust. We
have assumed a variable heat power $\Lh(t)$ and
calculated the variations of thermal surface luminosity $\Ls(t)$ or,
equivalently, of the surface temperature $\Ts(t)$ of the star to see
if these variations are observable.

Previously, we studied stationary heaters
\citep{2006Kam,2007Kam,2009HEAT,2014HEAT} where $\Lh$ and $\Ls$
were almost independent of time. Now we focus on variable heaters
which produce internal heat outbursts $\Lh(t)$ and 
corresponding surface outbursts $\Ls(t)$.
Recently,
we have investigated the problem \citep{2017Ch} 
neglecting superfluidity of neutrons in the inner 
crust. In
this paper, we include the effects of crustal superfluidity;
preliminary results have been outlined in \citet{JPCS}.

For illustration, we have taken two neutron star models, 
with $M=$1.4 and 1.85 \Msun\ and with the BSk21
EOS of nucleon matter in the star's core (Section
\ref{sec:level2}). We have considered three models
for singlet-state neutron superfluidity in the crust (Section
\ref{sec:level3}), and varied the heater's parameters. 

In Section \ref{sec:level4} we analyze relatively thin heaters. 
The results resemble those without superfluidity \citep{2017Ch},
but there are important differences. The common features
are like this. Only a small
amount of heat is emitted by photons through the surface. The
surface outburst $\Ls(t)$ is delayed and broadened with respect to
the internal outburst $\Lh(t)$, as a result of a finite heat
diffusion time $t_{\rm diff}$ from the heater to the surface. 
Also, we have confirmed the previous results  by other authors
(e.g., \citealt{2007Shternin,2009BC,2014HOMAN,2018Aql})
that the crustal superfluidity can greatly affect the
surface emission $\Ls(t)$ by reducing the heat capacity of free
neutrons and shortening $t_{\rm diff}$ in the inner superfluid crust
(provided the heater is there);
this makes the internal heater's activity more
visible at the surface.

In Sections \ref{sec:level6} and \ref{sec:relax},
we study outbursts produced by the deep crustal heating
\citep{2003HZ,1998BBR}  in neutron stars which enter
quasi-persistent LMXBs. Such an outburst is generated in a wide
crustal layer. It operates during accretion
stages and affects subsequent stages of long-term relaxation of
neutron stars in quiescence. We mainly focus on the late
stages of long-term relaxation. They can be characterized by
an effective relaxation time $\overline{\tau}$ and a longer but finite
duration time $t_{\rm rel}$ of the entire relaxation stage. 

The long-term relaxation depends on crustal superfluidity and
on the maximum density $\rhoacc$ to which the accreted matter sinks
in the crust. As long as the accreted matter is placed in the outer crust, 
the effect of superfluidity on  $\overline{\tau}$ and
$t_{\rm rel}$
is weak. If, however, the accreted matter penetrates into the inner
crust,  superfluidity becomes
very important. It reduces $\overline{\tau}$ and
$t_{\rm rel}$ in such a way as if the accreted matter occupies 
the outer crust alone.  Were the inner crust accreted but
non-superfluid, $\overline{\tau}$ and
$t_{\rm rel}$ would be a few times larger.

However, our analysis is too far from being completed. It would
be desirable to calculate a grid of models, particularly
for different $M$, $\dot{M}$, $\rhoacc$ and different models for crustal
superfluidity. While doing so, it would be good to use different
models for neutrino emission rates, heat capacities and thermal
conductivities in the accreted and non-accreted matter (whereas, for simplicity, we
have taken microphysics of the ground-state matter
throughout the entire crust, except for the heat energy release, Fig.\ \ref{fig:Qrho}). 

These results could be combined with the cooling theories which 
account for the effects of superfluidity in neutron star cores, for
instance, using the technique of \citet{2015Ofe,2017Ofe}. In
addition, in would be worthwhile to include proper models for heat
blanketing envelopes which strongly affect $\Ts(t)$ 
(e.g., \citealt{2016BPY}).

The obtained results can also be used to interpret the observations
of afterburst relaxation in magnetars (e.g.,
\citealt{MAGNETARS,2016LLB,2016BL,2017KB,2018MagnetarOutburs}) which may contain
variable  heaters distributed in their crusts. The afterburst relaxation
can have much in common with that in LMXBs (e.g. \citealt{2012AnKas}).
Since the magnetars possess strong magnetic fields of unknown
geometry and their heating mechanism is still not clear, it would
be a huge job to include the magnetic effects properly in calculations. 
For simplicity, these effects (discussed, e.g., in
\citealt{2003POT,PPP2015}) have been neglected in the present
study.

In any case, all these unsolved problems are beyond the scope of
this paper.

\acknowledgements
We are grateful to Sergei Balashev, Peter Shternin and Alexander Potekhin for a strong
and constructive criticism.
The work  was supported  by the Russian Science
Foundation (grant 14-12-00316).

\bibliographystyle{spr-mp-nameyear-cnd}


\end{document}